\documentclass[iopams,floats, floatfix, twocolumn,superscriptaddress, nofootinbib, showpac]{iopart}

\usepackage{graphicx}
\usepackage{xspace} 
\usepackage[usenames,dvipsnames]{color}
\usepackage{dcolumn}
\usepackage{bm}
\usepackage{iopams}
\usepackage{hyperref}
\usepackage{mathrsfs}

\usepackage{ulem}
\newcommand{\Caltech}{\address{Theoretical Astrophysics 350-17,
    California Institute of Technology, Pasadena, CA 91125}}

\newcommand{\CITA}{\address{Canadian Institute for Theoretical Astrophysics,
    University~of~Toronto, Toronto, Ontario M5S 3H8, Canada}}
\newcommand{\JPL}{\address{Jet Propulsion Laboratory,  California
    Institute of Technology, 4800 Oak Grove Drive, Pasadena, California 91109}}

\definecolor{darkgreen}{rgb}{0.2,0.7,0.2}

\newcommand{\he}{h_{\rm exact}}   
\newcommand{\hNR}{h_{\rm NR}}    
\newcommand{\hPN}{h_{\rm PN}}    
\newcommand{\hH}{h_{\rm H}}      
\newcommand{\NWIP}[2]{ \langle{#1},{#2}\rangle }    
\newcommand{\maxMM}{\varepsilon_{\rm max}}         
\newcommand{\SNR}{\rho}
\newcommand{\norm}[1]{\|#1\|}
\newcommand{\Q}{\norm{\delta h}/\norm{h}}

\begin{document}

\title[Suitability of PN/NR hybrid waveforms for
gravitational wave detectors]{Suitability of post-Newtonian/numerical-relativity hybrid waveforms for
gravitational wave detectors}

\author{Ilana MacDonald}\CITA
\author{Samaya Nissanke} \JPL \Caltech \CITA
\author{Harald P. Pfeiffer}\CITA

\begin{abstract}
  This article presents a study of the sufficient accuracy of
  post-Newtonian and numerical relativity waveforms for the most
  demanding usage case: parameter estimation of strong sources in
  advanced gravitational wave detectors.  For black hole binaries,
  these detectors require accurate waveform models which can be
  constructed by fusing an analytical post-Newtonian inspiral waveform
  with a numerical relativity merger-ringdown waveform.  We perform a
  comprehensive analysis of errors that enter such ``hybrid
  waveforms''.  We find that the post-Newtonian waveform must be
  aligned with the numerical relativity waveform to exquisite
  accuracy, about 1/100 of a gravitational wave cycle.  Phase errors
  in the inspiral phase of the numerical relativity simulation must be
  controlled to $\lesssim 0.1$rad.  
(These numbers apply to moderately optimistic estimates about 
  the number of GW sources; exceptionally strong signals require even 
  smaller errors.)
  The dominant source of error
  arises from the inaccuracy of the investigated 
  post-Newtonian
  Taylor-approximants.  Using our error criterium, even at 3.5-th post-Newtonian order,
  hybridization has to be performed significantly before the
  start of the longest currently available numerical waveforms which cover 30
  gravitational wave cycles. 
The current investigation is limited to the equal-mass,
  zero-spin case and does not take into account calibration
  errors of the gravitational wave detectors.  
\end{abstract}

\date{\today}


\maketitle

\section{Introduction}
\label{sec:intro}

Coalescing black-hole binaries are amongst the most promising sources
for the current and future gravitational wave (GW) detectors such as
LIGO, Virgo, LCGT~\cite{Barish:1999,Sigg:2008, Acernese:2008,Kuroda:2010} and LISA~\cite{lisa, lisa98, Jennrich:2009}. After
several years of instrument upgrades, the next
generation of ground-based interferometers should be operational
within five years ($\sim$ 2015-2016). At advanced sensitivities, currently predicted event rates
for stellar mass binary black holes (BBH) with total masses $<100 M_{\odot}$ range from
$0.4$ to $1000$ per year (with 20 being the ``realistic'' number
given in~\cite{AbadieLSC:2010}) detectable up to distances of
$\sim$several Gpc. (Henceforth, for simplicity, we consider LIGO to be representative of all ground-based GW detectors.) In contrast,
LISA will be sensitive to massive BBH systems with individual
masses of  $10^4-10^7 M_{\odot}$ up to redshifts
of $\sim$20. 

In order to detect GWs and derive the emitting sources' physical properties,
accurate source modeling of predicted GWs is required
in the form of a vast family of waveform templates. Analytical weak-field
approximation methods such as the post-Newtonian (PN) expansion in
general relativity accurately describe their inspiral prior to merger, whereas numerical
relativity (NR) is used to
model the merger of the two bodies in the strong--field
regime. Both techniques have come to fruition over the last decade,
e.g. the
reviews~\cite{Blanchet2006,Pretorius2007a,Hinder:2010vn,Centrella:2010}. Because of
the computational cost of numerical simulations ($\sim 100,000$
CPU-hours for long inspirals of generic precessing systems), current NR simulations cover
$\lesssim 15$ orbits.  These NR waveforms are then fused together with a
waveform covering the earlier stage of the inspiral, calculated using high order PN
expansions.  Such fused waveforms are referred to as \emph{hybrid} waveforms.  

Hybrids play an important role in the construction of PN inspired
phenomenological~\cite{Ajith:2008b,Santamaria:2010yb} waveforms, which are intended for event detection. In addition, hybrids may guide the construction of                                              
easy-to-implement waveforms that are being developed for measurement purposes. Within the Ninja project~\cite{NinjaWebPage,ninjashort}, moreover, GW data analysts search for hybrid
waveforms embedded in detector noise of
ground-based detectors to investigate efficiency of GW data-analysis
pipelines. Finally, the Numerical Relativity and Analytical
Relativity collaboration (NRAR)~\cite{NRARwebsite} aims to construct
analytical waveform models that  span the entire parameter
space of spins and mass ratios.  Such models
then allow us to compute waveforms  
with little computational expense for any choice of parameters.  Some of
the waveform models pursued in this collaboration might involve hybrid waveforms.  All these applications rest on sufficiently accurate hybrid waveforms, motivating our study.

Besides hybrid waveforms, analytical waveform models based on the effective-one-body (EOB)
formalism are common, e.g.~\cite{Buonanno2007,Buonanno:2009qa}. EOB waveforms
are a resummed extension of PN approximants, e.g.~\cite{Buonanno99}.
In this approach, one fits the EOB model directly against NR
waveforms, without the intermediate step of constructing hybrid
waveforms.  Such fits, too, will be sensitive to errors in the NR
waveforms.  Conclusions about required accuracy of NR
waveforms for hybrid waveforms may inform required
accuracies for the NR waveforms that are used when
fitting EOB models.

Both detection and measurement require hybrid waveforms to be of
sufficient accuracy, to avoid a negative impact on the signal
processing in noisy detector data. In conjunction with earlier
work~\cite{Miller2005}, Lindblom {\sl et al.} \cite{Lindblom2008} laid the foundations for accuracy standards for model
waveforms. The formalism is applicable to all model waveforms (including
PN only, NR only, and hybrids). In addition, though not considered in
this work, Refs.~\cite{Lindblom2009a,Lindblom2009b,Lindblom:2010mh}
extend the original model waveform analysis of Ref.~\cite{Lindblom2008} by incorporating
instrument calibration errors and deriving more refined accuracy requirements
within the time domain. 

Based on Ref.~\cite{Lindblom2008} and complementary to recent
investigations \cite{Buonanno:2009,Santamaria:2010yb,Hannam:2010,Damour:2010}, we implement waveform accuracy requirements that
preserve detection efficiency and parameter estimation performance. We
place bounds on the {\em norm} of the {\em modeling error} $\delta
h_{\mathrm{model}} = h_{\mathrm{model}}- h_{\mathrm{exact}}$, where
$h_{\mathrm{exact}}$ is assumed to be some exact waveform (a
solution of Einstein's equations at infinite precision), and
$h_{\mathrm{model}}$ is
an approximate model waveform. Specifically, the error measure
$\Q$ translates to a normalized noise-weighted cross-correlation of
the modeling error $\delta
h_{\mathrm model}$ with itself, and is thus instrument-dependent, but distance-independent. 

Our work computes this
quantitative \emph{accuracy
measure} in order to assess different sources of
error arising from hybrid waveforms for non-spinning, equal-mass binary black holes.  
Rather than emphasizing the immediate aim of event detection, we take a long-term perspective, and
  consider the ultimate accuracy requirements necessary for data-analysis
  once GW detections have become routine. The
  emphasis lies, therefore, on the optimal extraction of source characteristics from
  the observed waveforms. Due to the very nature of their
   construction, hybrids suffer
    from sources of error such as the type and location of matching the PN
and NR waveforms, and the PN and NR waveforms themselves. For
instance, PN methods rely on an expansion in powers of $v/c$ (where
$v$ is the BBH's typical orbital velocity and $c$ is the speed of
light)\footnote{The notation 1PN corresponds to the formal $\sim 1/c^2$ level in a post-Newtonian expansion with respect to 
the Newtonian acceleration and gravitational wave flux (where $c$ is the speed of light).}, and result in an inspiral waveform with both PN amplitude
and PN phase corrections. In addition, several different PN inspiral waveforms
 may be computed at a particular PN order because of differences when
truncating Taylor expansions (commonly referred to as Taylor T1,
Taylor T3, Taylor F2, etc., see e.g. Ref.~\cite{Damour:2000zb}).

An immediate question that arises when constructing hybrids is where matching
between their PN and NR waveforms should occur. This is of practical
importance because of the computational expense incurred in producing
NR waveforms.  Stitching PN and NR waveforms as late as
possible minimizes computational cost. However, PN waveforms, derived using
approximate perturbative methods in general relativity,
become increasingly inaccurate as merger is approached. We investigate here where best to match PN and NR waveforms. Other possible
sources of error, all defined below and explored
here within a unified framework, include numerical resolution, the extraction radius of NR
waveforms, the accuracy and approximant of the PN waveform, the matching width, and the effect of
error within the matching procedure itself. We also examine quantitatively how
detection and measurement accuracies vary using different ground-based
instrument noise curves, all of which have different regions of sensitivity.

This work thus provides a comprehensive quantitative measure of error for
waveforms from several sources. The paper
is organized as follows:  Sec.~\ref{sec:prev_work} describes related earlier work and Sec.~\ref{sec:notation} introduces the notation
used in this paper.  Sec.~\ref{sec:QuantifyingErrors} describes possible errors in hybrid waveforms and how to quantify them.  Sec.~\ref{sec:methodology} outlines 
our methodology: we begin with an explanation of the PN and NR waveforms in Secs.~\ref{sec:PNwaveforms} 
and~\ref{sec:NRwaveforms}, of the hybridization process in
Sec.~\ref{sec:hybridconstruction}, and of the computation of our error measures in Sec.~\ref{sec:Qcalc}. In
Sec.~\ref{sec:Results}, we present a detailed 
study of the different sources of error in hybrid waveforms. Sec.~\ref{sec:Discussion} presents our conclusions.

\subsection{Previous work }
\label{sec:prev_work}

Earlier works have examined similar aspects of detection
and measurement errors of hybrid waveforms in the case of Initial and
Advanced LIGO. Although the literature review presented here is far
from complete, we chronologically outline four works which follow methodologies
closely related to our study. As discussed in detail in Sec.~\ref{sec:QuantifyingErrors}, quantities in these works such as overlap, mismatch,
faithfulness, effectualness and inaccuracy are related to our error
measure $\Q$. Throughout the rest of the paper, we will
return to them to compare our results with their findings.

First, Buonanno {\sl et al.}~\cite{Buonanno:2009} compute overlaps between different time- and
frequency-domain PN waveforms and EOB waveforms for four 
non-spinning binaries with specific masses (the systems are (1.38,1.42)
$M_{\odot}$, (9.5,10.5) $M_{\odot}$, (10,1.4) $M_{\odot}$ and
(4.8,5.2) $M_{\odot}$). The overlaps are
maximized either over time- and phase-shifts (faithfulness), or over time- and phase-shifts and
masses (effectualness). Ref.~\cite{Buonanno:2009} does not consider hybrid or NR
waveforms {\sl per se}. Instead, it uses EOB waveforms
 that were
\emph{calibrated} with NR simulations. In the case of Initial, Enhanced
and Advanced LIGO, Ref.~\cite{Buonanno:2009}
concludes that the majority of Taylor approximants (excluding TaylorT3 and TaylorEt) are
sufficient for detection purposes for masses $< 12 M_{\odot}$
whilst
advocating the use of EOB for masses $>12 M_{\odot}$.

Second, whilst constructing a new set of phenomenological waveforms
for aligned and anti-aligned spinning BBHs, Santamar\'ia {\sl et al.}~\cite{Santamaria:2010yb} compute the normalized distance
squared $(\Q)^2$  (a quantity defined as the \emph{inaccuracy}, see
Ref.~\cite{Damour:2010}) for hybrids constructed by matching PN and NR
waveforms in the \emph{frequency} domain.  This differs from our study,
where matching proceeds only in the \emph{time}
domain. The PN waveforms comprise Taylor T1, Taylor
T4, or Taylor F2, and the NR
waveform in the non-spinning case is an earlier version
of the {\tt SpEC} waveform used here.  Santamar\'ia {\em et
  al.}~\cite{Santamaria:2010yb} determine the matching region itself by means of a least-squares fit between PN and NR waveform.  This results in a matching region which extends very close to merger, $M\omega \in
[0.063,0.126]$. Quantifiable error sources are given
by numerical resolution and by the
differences between Taylor T1, Taylor T4 or Taylor F2 hybrids. Ref.~\cite{Santamaria:2010yb} concludes that the PN truncation error, represented by the difference between different PN approximants, provides the
largest source of error. 

Third, and most similar to the approach followed here, Hannam {\em et
  al.}~\cite{Hannam:2010} aim to determine the necessary length of NR
waveforms for the construction of hybrid
waveforms. Ref.~\cite{Hannam:2010} considers both the mismatch
(minimized over time and phase-shifts with or without
  mass-optimization), equivalent to one minus the overlap, and
the inaccuracy,  $(\Q)^2$. As in Ref.~\cite{Santamaria:2010yb},
Ref.~\cite{Hannam:2010} constructs hybrid waveforms for aligned- and
anti aligned-spinning and non-spinning equal- and unequal-mass BBHs. Focusing on the
non-spinning equal-mass case, they use PN Taylor T1 and Taylor T4
approximants, and an earlier variant of the NR waveform used
here. Ref.~\cite{Hannam:2010} computes quantitative error measures
by comparing hybrids constructed from Taylor T1 and Taylor T4 with
different matching frequencies between $0.045$ and $0.09$, using an
Advanced LIGO noise curve with a low-frequency cut-off at
20Hz. Ref.~\cite{Hannam:2010} focuses on the near-term goal of event
detection, and therefore places significantly weaker accuracy
constraints on the hybrid waveforms.  Nevertheless,
Ref.~\cite{Hannam:2010} concludes that PN variants and PN order contribute the largest sources of error. 

Finally, Damour {\sl et al.}~\cite{Damour:2010} compute the
effectualness and inaccuracy between PN, phenomenological and EOB
waveforms for non-spinning unequal-mass systems. The PN waveform used
is Taylor F2, a frequency domain approximant also employed in
Santamar\'ia {\sl et al.} ~\cite{Santamaria:2010yb}. Similar to Ref.~\cite{Buonanno:2009}, they
do not consider NR waveforms and use an analytical Advanced LIGO noise
curve~\cite{Ajith:2008b}. Indeed, although similar in approach,
Ref.~\cite{Damour:2010} investigates different sources of errors in hybrid
and EOBNR waveforms than the present work. Ref.~\cite{Damour:2010} finds that
frequency-domain PN hybrids and phenomenological waveforms, unlike
their EOBNR counterparts, fail to satisfy the
minimal required accuracy standard for total mass $M < 75 M_{\odot}$, even for moderately low signal-to-noise ratio.

Shortly after the present article was submitted, a further study was performed by Boyle~\cite{Boyle:2011dy}.  This reference studies length requirements of Taylor-PN approximants hybridized with EOB waveforms.

Our work explores sources of errors in hybrid
waveforms, aspects of which were investigated by these earlier works~\cite{Buonanno:2009,Santamaria:2010yb,Hannam:2010,Damour:2010}, within a
uniform framework. In agreement with others, we find that the dominant
source of error is due to the uncertainty between PN Taylor-approximants.  Indeed,
using our error criteria, even at the highest available PN accuracy, we show that NR
waveforms must be longer than the currently available 30 GW cycles
when constructing hybrids with Taylor PN approximants in the non-spinning equal mass BBH case.

\subsection{Notation
\label{sec:notation}}

Throughout the paper, we use the following notation:

\begin{itemize}

\item $\he$ represents the {\em exact} waveform.
\item $\hPN$ and $\hNR$ denote the {\em
    post-Newtonian} and  {\em numerical relativity} waveforms respectively.
\item $\hH$ specifies a {\em hybrid} waveform. 
\item $\rho$ refers to the signal-to-noise ratio (SNR).
\item $\Q$ is our error measurement of the difference between a reference and trial hybrid.
\item $\mathcal{O}$ is the overlap between two waveforms.
\item $\mathcal{M}$ is the mismatch between two waveforms, defined as $1-\mathcal{O}$.
\item $\rho_{\rm eff}$ is the SNR divided by a detection safety factor
  $\epsilon$, and corrected for a network of detectors.  Its inverse $1/\rho_{\rm eff}$ is the upper limit for our error calculations.
\end{itemize}

\section{Quantifying errors}
\label{sec:QuantifyingErrors}

As mentioned in the introduction, hybrid waveforms are susceptible
to a wide range of possible errors.  We consider evolution of a binary black hole system with
certain given intrinsic parameters like mass ratio, spins, and orbital
elements.  If
one were able to solve Einstein's equations to infinite precision, one
would obtain the {\em exact} waveform $\he=\he(t)$ emitted by this
system.  For simplicity, we will only consider the dominant $l=2$, $m=2$ of the gravitational radiation. 

Numerical relativity computes an approximation to $\he$, which we
shall denote $\hNR$.  The NR waveform $\hNR$ may differ from the exact
waveform $\he$ for a variety of reasons, among
them:
\begin{enumerate}
\item  {\em Truncation error} due to finite numerical resolution.
  As numerical resolution is increased, the solution will converge to the
  solution of the underlying continuous initial-boundary-value problem.
\item  This continuous initial-boundary-value problem may differ from the
  realistic physical system, for instance, due to imperfect outer
  boundary conditions.
\item Gravitational radiation extracted at a finite radius may differ
  from the real waveform at future null infinity, due to
  ambiguities in the definition of gravitational radiation at finite
  radius, and imperfect wave extraction algorithms.
\item Initial data for the numerical simulation may not precisely
  correspond to the desired situation; for instance, the orbital
  eccentricity of the simulation might differ slightly from the
  desired value.
\end{enumerate}
This list is incomplete.  We further emphasize that many possible
sources of error for $\hNR$ do {\em not} converge away with increased numerical
resolution, such as, for example, the last three items in the list above.

The most striking difference between $\hNR$ and $\he$ is the length.
The exact waveform is infinitely long, with an infinite number of
gravitational wave cycles before merger.  In contrast, the NR waveform has a 
finite number of cycles.  Therefore, it is customary to
attach a PN inspiral $\hPN$ at the beginning of $\hNR$,
resulting in a {\em hybrid} waveform $\hH$ of sufficient length
to encompass the entire sensitivity band of the detector within a desired mass-range. 

Hybridization introduces several new potential sources of difference between
$\he$ and the hybrid waveform $\hH$:
\begin{enumerate}
\setcounter{enumi}{4} 
\item PN theory is in itself only an approximation to full
  general relativity; hence $\hPN$ will differ from $\he$ by some
  amount, which increases with decreasing separation of the binary.
  The accuracy of $\hPN$ also depends on the PN order of the approximant.
\item \label{item:PN-vs-NR-parameters}One must identify and match parameters
  like masses and spins in the PN formalism with those in
  the numerical simulation\footnote{This point was discussed in great
    detail in~\cite{Santamaria:2010yb}; here we simply choose masses
    and spins in the PN waveform to be equal to those in the NR
    simulation.}.
\item The matching between $\hPN$ and $\hNR$ is very sensitive to
  certain seemingly small errors in $\hPN$ or $\hNR$.
\end{enumerate}

To illustrate possible issues related to the matching, consider a
particularly simple hybridization procedure.  This procedure
chooses a reference frequency $x_{0}$ and time-shifts $\hNR$ and
$\hPN$ such that both waveforms pass through this frequency at the
same time.  Here, we parameterize frequency by the
PN-frequency parameter~\cite{Blanchet2006} $x=(M\Omega)^{2/3}$, where
$M$ is the total mass of the binary and $\Omega$ its orbital
frequency.  The sensitivity arises because the inspiral rate $\dot{x}$
decays very rapidly with decreased orbital frequency (i.e. increased
separation), with $\dot{x}\propto x^{5}$ during the inspiral at lowest
PN order.  Therefore, a small frequency-error $\delta x$ in
one of the two waveforms that are to be matched results in an error
\begin{equation}\label{eq:dt-from-dx}
\delta t = \frac{\delta x}{\dot{x}}\propto \frac{\delta x}{x^5} 
\end{equation}
in the time-offset necessary to align the two waveforms with each
other.  The term $x^{-5}$ causes the sensitivity of the
matching between PN waveform and NR waveform.  Matching typically
occurs at frequencies $x\sim 0.1$. Whether or not $\delta
t$ increases or decreases as one pushes the matching to lower frequency depends on how
quickly $\delta x$ decreases with earlier matching.  Let us
consider the error due to the (unknown) 4PN term contribution.  As a
function of time-to-coalescence $\Theta$, the 4PN term contributes 
an amount
\begin{equation}\label{eq:x4PN}
\delta x_{\rm 4PN}\propto \Theta^{-5/4}\propto x^5
\end{equation}
to the frequency.  The first proportionality arises from the PN
expansion of $x$ as a function of $\Theta$, (cf.~\cite{Blanchet2006}),
whereas the second proportionality arises from the  Newtonian relation
$x\propto \Theta^{-1/4}$.  Substituting Eq.~(\ref{eq:x4PN}) into
Eq.~(\ref{eq:dt-from-dx}), we see that the unknown 4PN coefficient
contributes an error $\delta t$ {\em independent} of the matching
frequency. 

As a second example, let us consider the orbital phase:  the (unknown) 4PN term of the orbital
phase, expressed as a function of frequency $x$ is 
\begin{equation}
\Phi_{\rm 4PN}\propto x^{3/2}\propto \Theta^{-3/8}.
\end{equation}
The impact of unknown higher-order PN terms on the orbital phase,
therefore, decays with smaller frequency $x$.  However, this decrease
is {\em very slow}: To reduce the effect of $\Phi_{\rm 4PN}$ by a
factor of 2, one has to terminate the PN waveform by a factor
$2^{8/3}\approx 6.4$ earlier in time, with a corresponding increase of
the length of the NR simulation which takes the binary to merger. 
Because of the high computational cost of an NR simulation, increasing
its length by large factors like 6.4 is hardly practical\footnote{Implicit time-stepping
  techniques~\cite{LauPfeiffer2008,Hennig:2008af,LauLovelacePfeiffer2011} are expected to
  reduce this limitation.}.  Furthermore, the numerical waveform
$\hNR$ also contributes potential issues for matching at
  earlier times.  Widely
separated binaries are particularly difficult to simulate numerically,
due to the small energy flux which drives the inspiral, and due to the
long wavelength of the gravitational radiation.  Moreover, the first
portion of numerical simulations tends to be noisy due to
junk radiation~\cite{Nissanke2006,hannamEtAl:2007,Lovelace2009,JohnsonMcDaniel:2009dq} arising from the initial conditions.

\begin{figure}
\centering 
\includegraphics[scale=0.39]{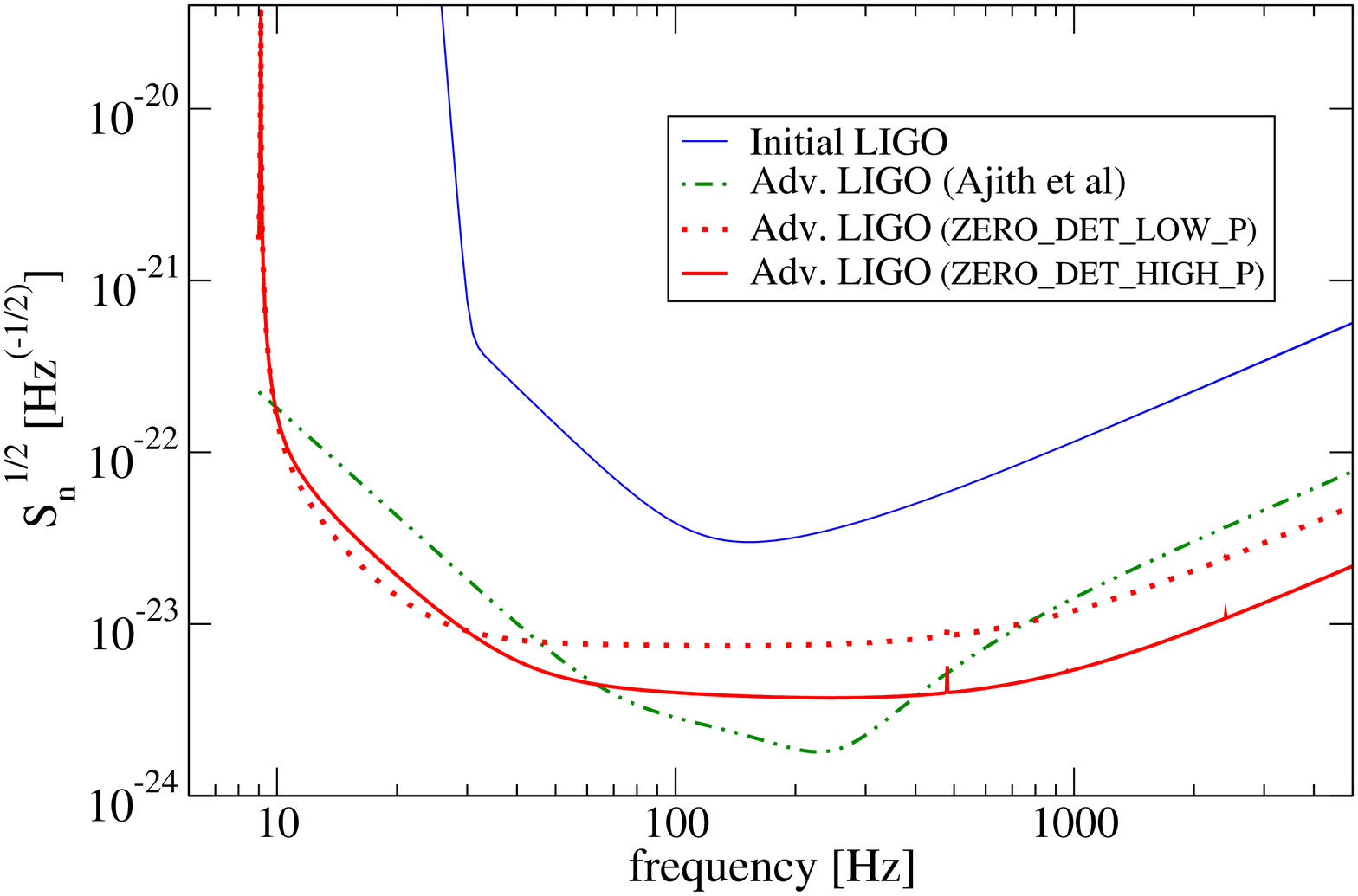}
\caption{\label{fig:aligonoise}
 LIGO noise curves used in our analysis.  We predominantly use
  {\tt ZERO\_DET\_HIGH\_P}, as predicted for Advanced
  LIGO~\cite{Shoemaker2009}. {\tt ZERO\_DET\_LOW\_P} is the Advanced LIGO sensitivity curve with a lower-powered laser.  In Sec.~\ref{sec:noise-curve}, we compare to two earlier
  noise curves, Initial LIGO, and an earlier Advanced LIGO noisecurve
  used by Ajith {\sl et al.}~\cite{Ajith:2008b}.  The wider bandwidth of
  {\tt ZERO\_DET\_HIGH\_P} places more stringent phase coherency on
  waveform templates.
}
\end{figure}

We see, therefore, that there are a wide variety of effects which
might cause $\hH$ to differ from the desired $\he$.  Whether such
differences are relevant will depend on the application of the
waveforms.  We focus here on applications to GW detection and, in
particular, measurement.  The sensitivity of GW detectors is frequency
dependent (see Fig.~\ref{fig:aligonoise}), and therefore, the importance of errors will
depend on their frequency dependence.  Because BBH waveforms are
invariant under a rescaling of mass, {\em one} PN-NR hybrid waveform
$\hH$ represents binaries of any total mass $M$ for a given mass ratio.  However, depending
on the mass $M$, different portions of $\hH$ are in the most sensitive
frequency range of the detector.  Furthermore, waveforms are typically
used in matched filtering, e.g. \cite{Finn:1992,CutlerFlanagan1994}, a technique that allows for time- and
phase-shifts of the gravitational waveform.  Thus, those components of
the difference $\hH-\he$ that can be represented by a time- or
phase-shift are not relevant.

Let us assume that the exact signal $\he$ and the computed hybrid
waveform $\hH$ are so close to each other that the Taylor-series
expansions used implicitly in the Fisher-matrix formalism are
valid\footnote{A commonly made assumption in gravitational wave
  data-analysis.}.  We denote
\begin{equation}
\delta h\equiv \hH-\he,
\end{equation}
where $\delta h$ is the difference between the hybrid and exact waveform when optimally aligned in phase and time.
To introduce some key concepts, we shall assume that we know $\delta h$.  Miller~\cite{Miller2005} and Lindblom {\sl et al.}~\cite{Lindblom2008} pointed out
that the key quantity for assessing the importance of errors is the
inner product $\norm{\delta h}^2\equiv \NWIP{\delta h}{\delta h}$, where
\begin{equation}\label{eq:NWIP}
\NWIP{g}{h}= 2 \int_0^{\infty} df \frac{\tilde{g}^*(f)\tilde{h}(f) +
\tilde{g}(f)\tilde{h}^*(f)}{S_n(f)} \;,
\end{equation}
with $\tilde{g}(f)$ and $\tilde{h}(f)$ the Fourier transforms of
two waveforms $g(t)$ and $h(t)$. $S_n(f)$ denotes the
(one-sided) power spectral density matrix,
\begin{equation}
\label{eq:sn_def}
S_n(f) = 2 \int_{-\infty}^{\infty} d \tau \, e^{2 \pi i f \tau}\,
C_n(\tau)\;,\qquad f>0,
\end{equation}
where $ C_n(\tau)$ is the noise correlation matrix for zero-mean,
stationary noise. The noise spectra employed in this paper are
  shown in Fig.~\ref{fig:aligonoise}.  We will generally focus on the
  noise curve {\tt ZERO\_DET\_HIGH\_P}, the highest-sensitivity design
  goal for Advanced LIGO~\cite{Shoemaker2009}. This configuration is sensitive over a
  wider bandwidth than the Initial LIGO, and earlier
   estimates for Advanced LIGO, used, for
    example, in Refs.~\cite{Ajith:2008b} and \cite{Damour:2010}.  Therefore, the PN
  inspiral and the NR late inspiral--merger--ringdown are
  simultaneously in band for a larger range of total binary masses than for previous noise curves. 
  For the earliest matching frequency considered here
  ($M\omega=0.038$), the matching region will be in Advanced LIGO's sensitivity band (defined here as 10Hz to 1000Hz) for total masses between about $1.2M_\odot$ and $125M_\odot$.

When
\begin{equation}\label{eq:ParamEstimation}
\norm{\delta h}<1,
\end{equation}
a detector with noise-spectrum $S_h(f)$ cannot experimentally
distinguish between the waveforms $\hH$ and $\he$. Therefore, $\hH$ is
a suitable substitute for $\he$ for parameter estimation in this
case.  
 The inequality~(\ref{eq:ParamEstimation}) is sufficient but not necessary.  For instance, if $\delta h$ is orthogonal to the signal manifold, its impact on parameter estimation is diminished.

Event detection genenerally places less restrictive demands on the
accuracy of the waveforms $\hH$.  A sufficient criterion is given by
Lindblom {\sl et al.}~\cite{Lindblom2008}, 
\begin{equation}\label{eq:Detection}
\norm{\delta h}<\sqrt{2 \maxMM}\;\SNR,
\end{equation}
where the signal-to-noise ratio (SNR)\, $\SNR\!=\!\norm{\he}$.
The dimensionless number $\maxMM$ depends on the reduction in SNR (and
thus event rate)  one is willing to
tolerate during event detection.  For typical template placement in
Initial LIGO data-analysis, Lindblom et
al. suggest $\maxMM=0.005$.  These numbers may have to be reconsidered
for Advanced LIGO data analysis~\cite{Owen:1996}, based on the cost of using a template
bank (which depends on the number of templates) relative to the cost of computing the templates in the first
place. 

For non-spinning equal mass BBHs, the waveforms $\he$ and $\hH$ depend trivially on two parameters: the luminosity distance and
total redshifted mass. The
luminosity distance to the source is introduced as an overall
scaling of the amplitude. For BBHs (as considered here),
the total redshifted mass comes into play only via trivial rescalings
of amplitude and time (i.e. frequency).  The dependence on distance
and mass carries forward into $\norm{\delta h}$, complicating
evaluation of Eqs.~(\ref{eq:ParamEstimation})
and~(\ref{eq:Detection}).  The distance dependence can be removed
through division by the signal-to-noise-ratio, i.e. by computing
\begin{equation}
\frac{\norm{\delta h}}{\SNR}.
\end{equation}
In terms of this quantity, the accuracy requirements become\footnote{For event detection, it is irrelevant which template matches
    a certain physical signal $\he$, and reliable estimates of
    detection sensitivities require mismatch calculations optimizing
    over physical parameters like total mass or mass-ratio.  We do not
    consider such optimizations in this paper, and therefore
    Eq.~(\ref{eq:dh-over-h-first}) gives only an approximate, conservative
    indication of requirements for event detection.  }
\begin{equation}\label{eq:dh-over-h-first}
\frac{\norm{\delta h}}{\norm{h}} <
\left\{\begin{array}{ll}
1/\rho_{\rm eff},&\mbox{parameter estimation}\\[0.6em]
\sqrt{2\maxMM}. &\mbox{event detection}
\end{array}
\right.
\end{equation}
The square of the left-hand-side of Eq.~(\ref{eq:dh-over-h-first}) is
called {\em inaccuracy functional} by Damour {\sl et
al.}~\cite{Damour:2010}.  Because of our emphasis on numerical errors,
we prefer to define this quantity without the square, so that $\Q$
will be proportional to the numerical errors.  This has the added
benefit that the right-hand side is proportional to the inverse of the
signal-to-noise ratio and not its inverse square.

For the parameter estimation limit in Eq.~(\ref{eq:dh-over-h-first}), we
  have replaced the single detector SNR $\rho$ by an {\em effective
    SNR} $\rho_{\rm eff}$\footnote{Lindblom {\sl et al.}'s original
    bound can be recovered by setting $\varepsilon=1$.}.  This
  effective SNR allows us to incorporate a safety factor
$\varepsilon<1$, as recently argued by Damour {\sl et
al.}~\cite{Damour:2010}.  The safety factor reduces the impact of
waveform errors $\delta h$ to a fraction $\varepsilon$ of the effects
of the detector
noise; Damour {\sl et al.}~\cite{Damour:2010} suggest $\varepsilon\sim
1/3 - 1/2$, in light of the fact that it is easier to
  calculate more accurate waveform templates than to enhance the
  sensitivity of the GW detectors. Furthermore, $\rho_{\rm eff}$ can
  absorb the impact of a {\em network} of detectors.  In a network of
  detectors with Gaussian, independent noise, the overall network SNR is the
  root-square-sum of the detector SNRs.  Thus, though GW detectors have
  non-Gaussian noise, one can still get a rough sense of the
  requirements for a coherent network of $N$ individual detectors by
  incorporating a factor $\sqrt{N}$ into $\rho_{\rm eff}$,
\begin{equation}\label{eq:rhoeff}
\rho_{\rm eff}=\varepsilon^{-1}\; \sqrt{N}\;\rho.
\end{equation}
Let us consider reasonable values for $\rho_{\rm eff}$. Our
  focus is on parameter
  estimation, i.e. we would like to obtain results that guarantee
  sufficiently accurate hybrids for most Advanced LIGO events.  In realistic
  scenarios, Advanced LIGO may be seeing 20 binary black holes per
  year~\cite{AbadieLSC:2010}.  Because LIGO is volume limited, most of
  these events will have an SNR close to the detection threshold, with
  the number of stronger events decaying like $\rho^{-3}$.  The
  strongest event in a year will have a SNR larger by a factor $\sim
  20^{1/3}=2.7$ than the detection threshold, placing it around SNR of
  $\approx 20$.  Including a safety factor $\varepsilon=1/2$, one
  arrives at $\rho_{\rm eff}=40$.  Accounting for 3 detectors
  increases this bound to $\rho_{\rm eff}=40\sqrt{3} \approx 70$ and
    for 5 detectors to $\rho_{\rm eff}=40\sqrt{5} \approx 90$.  Using the
  optimistic event rate estimate of 1000/yr~\cite{AbadieLSC:2010}
  instead of the realistic rate of 20/yr, increases the largest
  expected SNR's by a further factor $(1000/20)^{1/3}\approx 3.7$ to
  $\rho_{\rm eff}\approx 250$.  To cover this range of possibilities,
  we will indicate $\rho_{\rm eff}=40$ and
  $\rho_{\rm eff}=100$ in the plots below.  In addition, the event detection limit of
  Eq.~(\ref{eq:dh-over-h-first}), $\sqrt{2\varepsilon_{\rm max}}\approx
  0.1$, can be rewritten in terms of $\rho_{\rm eff}=10$.  We
  also indicate this bound in our figures. 
 
The dependence of $\norm{\delta h}$ and $\norm{\delta h}/\norm{h}$
  on the total mass can be taken into account by plotting $\Q$ as a function of total mass.
  Such a plot
  gives insight into suitability of $\hH$ for {\em both} event
  detection and parameter estimation; if $\norm{\delta h}/\norm{h}$ is
  below $\sqrt{2\maxMM}\sim 0.1$ for certain masses,
  criterion~(\ref{eq:Detection}) is satisfied, and $\hH$ is suitable
  for event detection in that mass-range.  The value of $\norm{\delta
    h}/\norm{h}$ as a function of mass gives the (inverse)
  signal-to-noise ratio, up to which $\hH$ is suitable for parameter
  estimation.  When a signal at a certain SNR $\SNR_{\rm obs}$ is
  observed, the plot allows one to verify whether
  $\norm{\delta h}/\norm{h}<\varepsilon/\SNR_{\rm obs}$.  If this
  inequality is satisfied, event characterization can proceed without
  delay; otherwise, more accurate waveforms $\hH$ need to be computed
  for optimal event characterization.

So far, we have assumed knowledge of the exact solution $\he$, which, in
practice, one does not know.  Instead, we will below compute
differences $\delta h=g-h$ between two hybrid waveforms $g$ and $h$, where we consider the ``superior'' of the two hybrid
waveforms as a substitute for $\he$.  Thus, we will usually consider
\begin{equation}\label{eq:dh-over-h}
\frac{\norm{\delta h}}{\norm{h}}.
\end{equation}
Replacing the denominator of Eq.~(\ref{eq:dh-over-h}) by $\norm{g}$,
or by $\norm{g}^{1/2}\norm{h}^{1/2}$ changes Eq.~(\ref{eq:dh-over-h})
only by terms of higher order in $\delta h$ and is therefore
irrelevant.  We will, therefore, consider Eq.~(\ref{eq:dh-over-h}) as a
measure of the error of the ``inferior'' of the two waveforms $g$,
$h$.  In doing so, it is important that one of the waveforms is
clearly more accurate than the other, otherwise
Eq.~(\ref{eq:dh-over-h}) would simply measure the difference between
two poor-quality waveforms (and such a difference can be arbitrarily small, if
the two poor-quality waveforms happen to be near each other).  We will discuss
this point in more detail below.

The error measure $\norm{\delta h}/\norm{h}$ can be expressed in terms
of the overlap $\mathcal{O}$ and the mismatch $\mathcal{M} = 1 -
\mathcal{O}$.  The overlap between two waveforms $g$ and $h$ is defined as
\begin{equation}
\mathcal{O} = \frac{\NWIP{g}{h}}{ \sqrt{\NWIP{g}{g}\NWIP{h}{h}}} 
= \frac{\langle h +\delta h\vert h \rangle}{ \sqrt{\langle h+\delta h \vert h+\delta h \rangle \langle h \vert h \rangle}}.
\end{equation}
By Taylor expanding the denominator, we find:
\begin{equation}
\mathcal{O} = 1 - \frac{1}{2}\, \frac{\norm{\delta h}^2}{\norm{h}^2} + \frac{1}{2}h_{\parallel}^2 +\mathscr {O}(\delta h^3),
\end{equation}
and
\begin{equation}
\label{eq:mismatch}
\frac{\norm{\delta h}}{\norm{h}} =
\sqrt{2{\cal M} + h_{\parallel}^2\;}+\mathscr {O}(\delta h^2).
\end{equation}
Here, $h_{\parallel}$ is the normalized projection of $\delta h$ onto $h$, 
\begin{equation}\label{eq:h-parallel}
h_{\parallel}=\frac{\NWIP{\delta h}{h}}{\NWIP{h}{h}}+\mathscr {O}(\delta h^2).
\end{equation}
Normalizing Eq.~(\ref{eq:h-parallel}) by $\norm{g}^2$ instead
  of $\norm{h}^2$ merely affects higher order terms.  The quantity
  $h_\parallel$ represents an overall rescaling between $h$ and $g$;
  if $g=(1+\gamma)h$ for some constant $\gamma$, then
  $h_\parallel=\gamma$.

Eq.~(\ref{eq:mismatch}) relates our error measure $\Q$ to the more widely used mismatch ${\cal M}$.  The major difference
  is that $\Q$ is also sensitive to an overall rescaling of the
  waveform, represented by $h_{\parallel}$, whereas ${\cal M}$ is not.
  Waveforms $h$ are high-dimensional vectors either in the time-, or
  frequency-domain, and the quantity $h_{\parallel}$ only measures one dimension of
  this high-dimensional vector space.  If one assumes errors that have
  components in many dimensions, and consequently neglects
  $h_\parallel$, then Eq.~(\ref{eq:mismatch}) simplifies to
\begin{equation}\label{eq:Q-missmatch-approx}
  \Q=\sqrt{2{\cal M}}.
\end{equation} 
This relationship has also been noted in~\cite{Lindblom2008,Santamaria:2010yb,Hannam:2010,McWilliams2010b}.

 When we compute $\Q$ in subsequent
  sections, we always introduce a relative time- and phase-shift
  between $g$ and $h$ in order to minimize $\Q$ and the mismatch ${\cal M}$,
  but we do not modify any other parameters of the waveform
  (such as the total mass).  Thus, our analysis tests faithfulness~\cite{Damour:2000zb}.
A more appropriate error limit for detection is the {\em effectualness}
where the error is minimized over the intrinsic parameters of the system, that is,
the total mass in an equal-mass, non-spinning binary.  Thus, our upper error limit
of $\rho_{\rm eff} = 10$ for detection is very conservative.

Finally, we note that the analysis presented here is not the first one
in this spirit.  Mismatches or $\Q$ between various pairs of waveforms have been calculated before, for instance
between purely numerical waveforms in the Samurai
project~\cite{Hannam:2009hh}, between analytical waveform models by
Damour {\sl et al.}~\cite{Damour:2010}, and also between hybrid
waveforms~\cite{Santamaria:2010yb,Hannam:2010}.  Our focus is a
comprehensive and unified study of many error sources that enter hybrid
waveforms, concentrating on parameter estimation.

\section{Methodology}
\label{sec:methodology}

\subsection{Post-Newtonian waveforms \label{sec:PNwaveforms}}

The PN approximation in General Relativity relies
critically on assumptions of weak field and of the source's internal slow motion. It is characterized by the PN parameter
$\epsilon \sim (v/c)^2 \sim (M/r)$, where $M$ is the characteristic
mass of the system and $v$ the magnitude of the relative velocity.
Construction of PN templates requires modeling both the
local conservative dynamics of the system and the generation of its
gravitational waves. The current state-of-the-art accuracy in both the
gravitational wave generation and the description of the system's
local dynamics is 3.5PN in the case of non-spinning quasi-circular
comparable mass inspiralling compact binaries (see
Blanchet~\cite{Blanchet2006} and references therein for a detailed review). In the case of spinning
inspiralling compact binaries, the amplitude-complete GW templates have been computed to
1.5PN order~\cite{Arun:2009}.  Following standard notation,
  $n$PN describes the $n$-th PN order in the phase, and ``PN
approximant $n/m$'' specifies the $n$-th and $m$-th PN corrections in the phase
and amplitude respectively.

Standard formulae exist for PN gravitational polarization waveforms
$h_+(t)$ and $h_\times(t)$ of quasi-circular orbits, which depend on
the source's location with respect to some observer.  Usually, NR waveforms are  decomposed into
spherical harmonics and often only the dominant (2,2) mode is
 considered. This is the case with the set of NR simulations we use, and consequently, we will only need the (2,2) mode of the PN approximant~\cite{Kidder:2007rt}:
\begin{eqnarray}
\fl h_{(2,2)} = & -2 \sqrt{\frac{\pi}{5}} \frac{G M}{c^2 R} e^{-2i \Phi} 
x \Bigg\{ 1 - \frac{373}{168} x + \left[ 2 \pi + 6 i
  \ln\Big(\frac{x}{x_0}\Big) \right] x^{3/2} -
\frac{62653}{24192} x^2\nonumber \\
\fl &  - \bigg[
  \frac{197}{42} \pi   + \frac{197 i}{14}
  \ln\Big(\frac{x}{x_0}\Big) + 6 i \bigg]
x^{5/2}
\label{eq:h22}
+ \bigg[
  \frac{43876092677}{1117670400} + \frac{99}{128} \pi ^2 
- \frac{428}{105} \ln{x} \nonumber\\ 
\fl & - \frac{856}{105} \gamma - \frac{1712}{105}
  \ln{2}
  - 18 \ln^2\Big(\frac{x}{x_0}\Big)
     + \frac{428}{105} i \pi + 12 i \pi
  \ln\Big(\frac{x}{x_0}\Big) \bigg] x^3
\Bigg\},
\end{eqnarray}
where the PN invariant velocity parameter is $x\!\equiv\!
\left( M \Omega (t) \right)^{2/3}$, $\Omega (t)$ is the orbital
frequency and $\Phi(t)\!\equiv\! \int^{t}_{0} \Omega (t') dt'$ is the
orbital phase. The parameter $\ln{x_0}\!\equiv\!\frac{11}{18} -
\frac{2}{3}\gamma + \frac{2}{3}\ln{\left(\frac{GM}{4bc^3}\right)}$ is
a constant frequency scale which depends on the freely specifiable
time parameter $b$. The latter characterizes the timescale at which
the gravitational wave tail contributions enters the polarization
waveforms. We choose $b=1$. 

Evaluation of (\ref{eq:h22}) requires expressions for $x(t)$ and
$\Phi(t)$. We will investigate the Taylor approximants TaylorT1, TaylorT3 and
TaylorT4 (as defined in~\cite{Boyle2007}  and references therein). 
They are all completed by solving the energy-balance equation:
\begin{equation}
\label{eq:EnergyBalance}
\frac{dE}{dt} = - {\cal L},
\end{equation}
where the GW flux at infinity ${\cal L}$ is balanced
by the change in the orbital binding energy $E(t)$ of the
binary.  The TaylorT approximants differ from each other by how higher-order PN terms are truncated in
$E(t)$ and ${\cal L}$ when implementing (\ref{eq:EnergyBalance}). In particular, TaylorT1 and TaylorT4 are solved by
numerically integrating coupled sets of ordinary differential
equations (see (35-36) and (45-46) in~\cite{Boyle2007}
respectively and~\cite{Damour:2000zb,Damour:2002kr}), whereas analytical expressions
exist for TaylorT3 (see (43)-(46) in~\cite{Boyle2007} and~\cite{Blanchet2006}). Many previous studies
  have established that for equal-mass, non-spinning binaries the
  TaylorT4-approximant happens to agree very closely with the full
  numerical solution of Einstein's
  equations~\cite{Baker2006e,Buonanno-Cook-Pretorius:2007,Boyle2007,Hannam2007}. There
  is, however, no {\it a priori} reason why TaylorT4 should perform
better than TaylorT1 or TaylorT3. So that our results are as representative as possible, we therefore
investigate several Taylor approximants (TaylorT1, T3, T4). By doing so, we
  achieve conservative estimates of how long NR simulations must be.
  We expect that our results are unaffected by accidental agreement of
  TaylorT4 for the specific configuration under study, so that our
  results will carry over to other parameter configurations where no
  exceptionally good PN approximant is known.

We start the PN waveforms at an initial orbital
    frequency of $M\Omega= 7.5\times 10^{-4}$, resulting in a waveform of
    duration $\sim 1.7\times 10^7M$ containing $\sim 6500$ GW cycles to
    merger.  The low starting frequency ensures an initial
    gravitational wave frequency of 10 Hz at $M=5M_\odot$, so that the
    constructed hybrid waveforms cover the entire frequency band of the
    Advanced LIGO detectors for $M\ge 5M_\odot$.

\subsection{Numerical Relativity Waveforms \label{sec:NRwaveforms}}

In this paper, we focus on equal-mass non-spinning BBHs.
We use a waveform generated with the {\em
  Spectral Einstein Code} {\tt SpEC}~\cite{SpECwebsite}.  The initial
data is identical to that used
in~\cite{Boyle2007,Scheel2008}, resulting in an
inspiral lasting about 16 orbits at an initial orbital eccentricity of
$\sim 5\times 10^{-5}$.  The techniques for the evolution are a
refinement of those presented in~\cite{Boyle2007,Scheel2008}.
Specifically, the gauge source functions are rolled off to zero early
in the evolution, so that the inspiral proceeds in the harmonic gauge.
The constraint-damping parameters are those described
in~\cite{Chu2009}, which reduce the impact of junk radiation.  The
merger was performed with a refinement of the techniques described
in~\cite{Szilagyi:2009qz}.

\subsection{Construction of hybrid waveform 
\label{sec:hybridconstruction}}

Construction of a hybrid waveform $\hH (t)$ from a PN waveform $\hPN(t)$
and NR waveform $\hNR(t)$ requires first that we align the waveforms
with a relative time- and phase-shift, and then join them
smoothly together. We begin by choosing a GW frequency interval of width $\delta\omega$ centered at frequency $\omega_m$,
\begin{equation}\label{eq:omega_match}
\omega_m-\frac{\delta\omega}{2} \le \omega \le \omega_m+\frac{\delta\omega}{2},
\end{equation}
in which we perform the matching between NR and PN.  During the
inspiral, the gravitational wave frequency $\omega(t)=d\phi(t)/dt$ is continuously increasing, so
(\ref{eq:omega_match}) translates into a time-interval $t_{\rm
  min}<t<t_{\rm max}$.  Here, $\phi(t)$ represents the phase of the
(2,2) mode of the gravitational radiation.

The PN waveform $\hPN(t;t_c,\Phi_c)$ incorporates
naturally the coalescence parameters $t_c$ and $\Phi_c$ representing the time and
phase freedom. We determine the parameters $t_c'$ and $\Phi_c'$ by
minimizing the GW phase difference in the matching interval $[t_{\rm min}, t_{\rm max}]$,
\begin{equation}
\label{eq:rms-phase-diff}
t_c,' \Phi_c' = \mathrm{ \mathop{minloc}_{t_c, \Phi_c}}\int^{t_{\rm max}}_{t_{\rm min}} \big(
  \phi_{\rm PN}(t;t_c,\Phi_c) - \phi_{\rm NR} (t) \big)^2 \rm{d}t.
\end{equation}
The hybrid waveform is then constructed in the form
\begin{equation}
\hH(t) \equiv \mathcal{F}(t) \hPN(t;t'_c,\Phi'_c) + \big[1- \mathcal{F}(t)\big]  \hNR (t), 
\end{equation}
where $\mathcal{F}(t)$ is a blending function defined as
\begin{eqnarray}
\mathcal{F}(t) \equiv 
\left\{
\begin{array}{ll}
  1, &  t < t_{\rm min} \\ 
 \frac{1}{2}\left(1+\cos\frac{\pi(t - t_{\rm min})}{t_{\rm max} - t_{\rm min}}\right),\quad\quad &  t_{\rm min}
  \leq t < t_{\rm max} \\ 
  0. & t\geq t_{\rm max}  
\end{array}
\right.
\end{eqnarray}

Several other hybrid procedures have been proposed. Similar
to our method, Hannam {\sl et al.}~and Ajith {\sl et
  al.}~\cite{Hannam:2010,Ajith:2008b} match in the time domain,
although they employ different blending functions (e.g., Eq.~(4.11) in~\cite{Ajith:2008b}). Both Santamar\'ia {\sl et al.}~\cite{Santamaria:2010yb} and Damour
{\sl et al.}~\cite{Damour:2010} match in the frequency domain using a $\chi^2$ fit,
and invoke different blending functions. Earlier hybrids were matched at a single
frequency~\cite{Boyle2007,Hannam2007}.  We have experimented with
  minimizing the phase and amplitude difference when determining
  time and phase offset. Specifically, we replaced the integrand
  of~(\ref{eq:rms-phase-diff}) by $|\hPN-\hNR|^2$ using the complex
  (2,2) modes.  We did not notice any significant change
  when using this different
  alignment scheme.

\subsection{Computation of $\Q$ \label{sec:Qcalc}}

We compute Fourier transforms of waveforms with standard Fast
Fourier Transforms, rescaling the geometric units in which the hybrid
waveforms are given to the desired physical mass.  In order to reduce
the Gibbs phenomenon, we multiply the time-domain waveform with a windowing function $w(t)$ before computing the Fourier transform.  We tried two windowing functions, the Hann (or cosine
squared) window~\cite{Blackman-Tukey} and the more recent Planck-taper window function~\cite{McKechan:2010kp}. Our results did not differ noticeably when
using the different windowing functions, and so the
Planck-taper window~\cite{McKechan:2010kp} is used since
it quickly goes to zero outside the window bounds: 
\begin{equation}
w(t) =
\left\{ \begin{array}{ll}
  0, &  t \leq t_{1} \\
  \left[e^{y(t)}+1\right]^{-1}, \qquad&  t_{1} < t < t_{2}\\
  1, &  t_{2} \leq t \leq t_{3} \\
  \left[e^{z(t)}+1\right]^{-1}, &  t_{3} < t < t_{4}\\
  0, &  t_{4} \leq t,
\end{array} 
\right.
\end{equation}
where $y(t) = (t_{2}\!-\!t_{1})/(t\!-\!t_{1}) +
(t_{2}\!-\!t_{1})/(t\!-\!t_2)$ and $z(t)
=-(t_{3}\!-\!t_{4})/(t\!-\!t_{3}) + (t_{3}\!-\!t_{4})/(t\!-\!t_4)$. The interval $[t_1, t_2]$ is chosen to cover the first five GW cycles of our hybrid waveform, whereas $[t_3,t_4]$ spans the very late ringdown.

Detector noise has a considerable effect
on $\NWIP{h}{g}$, since different parts of the waveform enter the
detector band for different masses. When evaluating the
  noise weighted inner products in Eq.~(\ref{eq:NWIP}), we use the
    noise curves shown in Fig.~\ref{fig:aligonoise}, with the default being the Advanced LIGO {\tt
          ZERO\_DET\_HIGH\_P} curve.

 Whenever we calculate $\Q$ between two hybrid waveforms, we time-- and phase--shift one of the two hybrid waveforms to maximize the overlap for each mass. We can write the Fourier transforms of the time- and phase-shifted waveform as follows:
\begin{equation}
\label{eq:phasetime}
\tilde{h}_{\Delta\phi, \Delta t} = (\tilde{h}_{c} \cos\Delta\phi + \tilde{h}_{s}\sin\Delta\phi) e^{2\pi i f \Delta t}.
\end{equation}
 The extremization over the
phase-shift $\Delta\phi$ can be  performed
analytically.  We define $a = \langle \tilde{h}_{c}, \tilde{g}
\rangle$ and $b = \langle \tilde{h}_{s}, \tilde{g} \rangle$, where
$\tilde{h}_{c}$ is the Fourier transform of the real part of the first
waveform, and $\tilde{h}_{s}$ is the Fourier transform of its
imaginary part.  Thus,
\begin{equation}
\Delta \phi = \tan^{-1}{\left( \frac{b}{a} \right)},
\end{equation}
which is the phase shift for maximum overlap if $[-\sin{\Delta\phi}
- \cos{\Delta\phi}] < 0$ (otherwise $\pi$ is added to the
phase). $\Delta t$ is found by using the Matlab minimization function
{\tt fminsearch}.

\section{Results \label{sec:Results}}

In the following error analysis, we focus primarily on the upper
limits of $\Q$ for a single detector, corresponding to $\rho_{\rm
  eff}$ of 10 and 40 for detection and parameter estimation
respectively. We do not discuss the upper error limit for multiple
detectors, $\rho_{\rm eff} \sim 100$, since none of our hybrid waveforms
meet this error criterion, with the possible exception of those
hybrids matched very early with a TaylorT4 waveform.

\subsection{Effect of error in hybridization}
\label{sec:41}

During the hybridization procedure, as outlined in Section
\ref{sec:hybridconstruction}, the PN waveform must be aligned with the
NR waveform by choosing $t_c$ and $\Phi_c$. The time shift $t_c$ is
difficult to determine, because a degeneracy between time and phase shift is only 
broken by the frequency evolution, which is very slow  during the inspiral, cf. Sec.~\ref{sec:QuantifyingErrors}.  On the other hand, minimization over the phase-shift $\Phi_c$ is trivial, because $\Phi_{\rm PN}(t; t_c, \Phi_c)-\Phi_{\rm NR}(t)$ in the integrand of Eq.~(\ref{eq:rms-phase-diff}) is linear in $\Phi_c$.
Therefore, we assume that errors in the determination of $\Phi_c$ are irrelevant,  and focus on the time-offset by investigating how accurately $t_c$ must be determined.  
For this, we first construct a hybrid using the correct least-squares fit, 
resulting in the correct $t'_c$.  We then construct a second hybrid by fixing
the time-offset to $t_c'+\Delta T_{\rm bias}$ and by performing the
minimization over $\Phi_c$ only.  We compute $\Q$ between these
two hybrids.  All least-squares fits are performed over a frequency
range of width $\delta\omega=0.1\omega_m$ centered at frequency
$M\omega_m=0.05$.

\begin{figure}
\centerline{\includegraphics[scale=0.37]{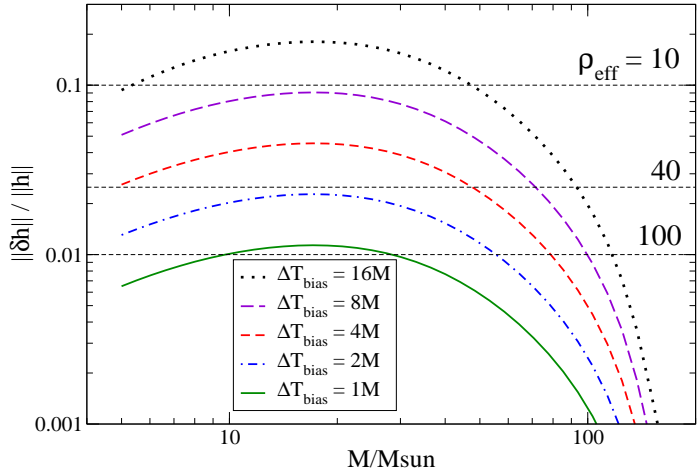}}
\caption{\label{fig:HybridErrors}Effect of hybridization
    errors.  We compare hybrid waveforms with alignment biased by
    $\Delta T_{\rm bias}$ to the correctly hybridized waveforms.
    (Matching frequency $M\omega_m = 0.050$, highest-resolution NR
    waveform, and TaylorT3 PN waveform).}
\end{figure}

Fig.~\ref{fig:HybridErrors} presents the results of this calculation. 
The error $\Q$ decreases linearly with $\Delta T_{\rm bias}$; this
linear dependence is one advantage of using $\Q$ over other measures, such as
its square or the mismatch.   Fig.~\ref{fig:HybridErrors}
  shows that even small errors in hybridization have a significant
  effect on the magnitude of $\Q$.  Time-shift errors larger than
  about 2M are unacceptable for parameter estimation at
  $\rho_{\rm eff}\gtrsim 40$.  In order to preserve some
  safety margin, we therefore conclude that the time-shift must be
  determined to better than $1M$. (Strictly speaking, this estimate $\Delta T\lesssim 1M$ applies only for matching at a
  frequency $M\omega_m\approx 0.05$; this analysis would have to be
  repeated when matching earlier or later).  This is a
surprisingly tight bound; $1M$ is only $\sim 1/2000$ of the
time-to-merger, and is only about $1\%$ of a gravitational wave
cycle.

Our primary goal is to utilize the  bound on $\Delta T_{\rm bias}$ when choosing the width of the matching interval in Sec.~\ref{sec:width}.  However, to place this requirement $\Delta T_{\rm bias}\lesssim 1M$ into
perspective, consider the TaylorT2 PN approximant, which
expresses time-to-merger as a power-series in orbital frequency.
The 3.5PN-order contribution to the coalescence time for an equal-mass binary is given
by (see \cite{Blanchet2006} and references therein);
\begin{equation}\label{eq:t_35PN}
t_{\rm 3.5PN}= \frac{5M}{64}\frac{571496}{3969}\pi (M\Omega)^{-1/3}\approx 96M,
\end{equation}
where we have used $M\Omega=M\omega_m/2=0.025$.  The contribution
$t_{\rm 3.5PN}$ is two orders of magnitude larger than the bound on
$\Delta T_{\rm bias}$.  Whilst it is difficult to translate time-offsets from
PN contributions, this large disagreement
immediately indicates that PN theory, even at 3.5PN
order, might not be sufficiently accurate.

We note that Hannam {\sl et al.}~\cite{Hannam:2010} also considered
hybridization errors; this work compared three different hybridization
schemes, and found that their difference is indistinguishable for
certain SNRs.  Our approach improves on this by explicitly giving conditions
on the accuracy of $t_c$ as in Fig.~\ref{fig:HybridErrors}.  This
measure of hybridization errors might seem extremely specific because we only 
consider the errors due to the determination of $t_c$, however these error
bounds apply to any time-domain hybridization procedure and to any effect that might give rise to an error in determination of $t_c$, like for instance a small eccentricity in the NR simulation.

\subsection{Width of hybridization interval}
\label{sec:width}

\begin{figure}
\centerline{\includegraphics[scale=0.53]{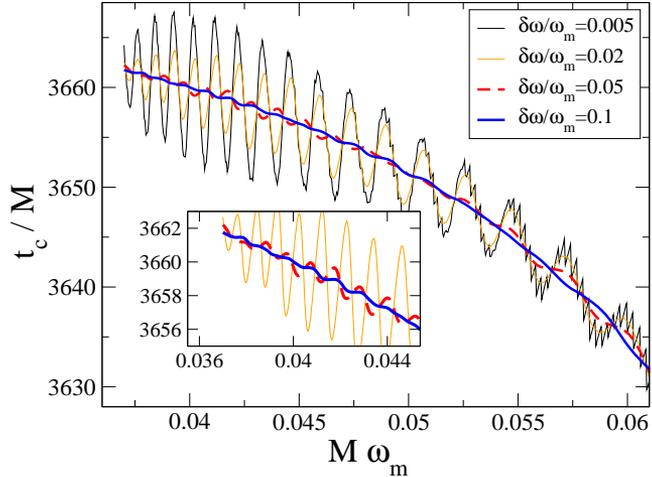}}
\caption{\label{fig:WidthOfMatchingWindow} Effect of the width of
  matching window $\delta\omega/\omega_m$ on the best-fit time-shift
  $t_c$. For matching windows covering $\lesssim 1$ GW period
  ($\delta\omega/\omega_m\lesssim 0.02$), $t_c$ is sensitive to
  small oscillations in the GW phase (thin lines).  For windows
  covering $\gtrsim 1$ GW period, this sensitivity rapidly decreases
  (thick lines, also enlarged in the inset).  }
\end{figure}

The PN time- and phase-parameters $t_c$ and $\Phi_c$ that align the
PN waveform with the NR waveform are determined by a least-squares
fit [see Eq.~(\ref{eq:rms-phase-diff})].  The width of the matching
interval $[t_{\rm min}, t_{\rm max}]$ influences the robustness of
this fit for two reasons:
\begin{enumerate}
\item A small change $\delta t_c$ in $t_c$ is nearly degenerate with a small
  change $\delta\Phi_c$ in $\Phi_c$, with $\delta\Phi_c\approx
  \omega_m\delta t_c/2$. (The factor $1/2$ arises because $\omega_m$ measures the frequency of the GW $l=2, m=2$ mode whereas $\Phi_c$ is the orbital phase). For circular Newtonian orbits, the
  degeneracy is perfect.  For the general relativistic case, the
  degeneracy is broken only by the {\em evolution} during the
  hybridization interval, i.e. by the change in frequency $\omega$.  A
  longer fitting interval encompasses more evolution, and so helps to
  break the degeneracy.
\item A longer fitting interval helps to ``average out'' undesirable features in the NR waveform.  Such features might be
  caused by numerical artefacts like junk radiation or a slight
  residual eccentricity.  In particular, matching over an interval much
  smaller than the orbital period will be susceptible to eccentricity
  effects, as discussed in Boyle {\sl et al.}~\cite{Boyle2007}.
\end{enumerate}
A robust fit Eq.~(\ref{eq:rms-phase-diff}) must be independent of small
changes to the fitting interval.  To test this, we compute many
independent least-squares fits, over slightly shifted fitting
intervals, parametrized by $\omega_m$, the frequency in the middle of
the fitting interval.  Fig.~\ref{fig:WidthOfMatchingWindow} plots the
resulting $t_c$ for several different widths $\delta\omega$ of
the matching interval. 

For small $\delta\omega$, this figure shows significant variations in
$t_c$.  These variations arise because the fitting interval
covers less than one GW period.  Therefore, the
fitting is sensitive to a small variation in the GW phase (of order
$0.001$ rad) present in the numerical data.  As
$\delta\omega/\omega\gtrsim 0.05$, this sensitivity rapidly decreases,
as the fitting interval now encompasses multiple cycles, averaging
over the small oscillations in the numerical waveform.  The remaining
overall trend ($t_c\approx 3660M$ to $t_c\approx 3630M$) is
explained by the increasing error in the PN waveform at higher
frequencies; the value $\sim 3650M$ arises because the time in the NR simulation is chosen such that $t=0$ at the beginning of the simulation, with merger around $t\sim 4000M$.  For $\delta\omega/\omega=0.1$ (solid blue curve),
residual oscillations in $t_c$ are below $1M$, as can be seen
from the inset of Fig.~\ref{fig:WidthOfMatchingWindow}.  We therefore
recommend a matching width of at least $\delta\omega/\omega=0.1$, and
adopt this value for the remainder of this paper.

To the lowest PN order, the time to merger is proportional to $M\omega^{-8/3}$. Therefore, a relative frequency width of 10\%
corresponds to the first {\em quarter} of the NR waveform.  The PN
waveform must be sufficiently accurate throughout the entire matching
interval, so this necessitates a NR waveform which extends
considerably into the regime where PN errors are small.

\subsection{Systematic errors in the NR waveform}
\label{sec:43}

Let us now examine two errors that arise during a numerical
simulation: numerical truncation errors and errors introduced by
gravitational wave extraction at finite radius.

\begin{figure}
\centerline{\includegraphics[scale=0.38]{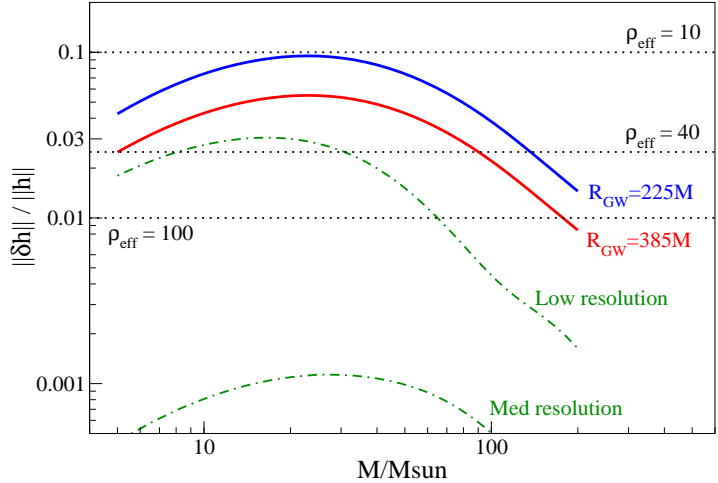}}
\caption{ \label{fig:Numerical_Resolution} Effect of numerical errors.
  Shown is $\Q$ between hybrid waveforms constructed with NR waveforms
  of different numerical resolution (dashed lines), and between hybrid
  waveforms of finite GW extraction radius $R_{\rm GW}$ vs. waveforms
  extrapolated to infinite extraction radius.  (All hybrids matched at
  $M\omega = 0.042$ to a TaylorT3 PN waveform). }
\end{figure}

We assess the truncation error by constructing hybrids from NR
waveforms at three different numerical resolutions (N4, N5 and N6 in
Ref.~\cite{Boyle2007,Scheel2008} with approximately $57^3$, $62^3$,
and $67^3$ grid points respectively).  These hybrids utilize a
TaylorT3 PN waveform and $M\omega_m=0.042$.  We then compute $\Q$
between low and medium resolution, and between medium and high resolution.
Because of the rapid convergence of the spectral methods used to
compute these numerical waveforms, the errors $\Q$ essentially
represent the error of the lower of the two resolutions involved.  The
results are plotted as the dashed-dotted lines in
Fig.~\ref{fig:Numerical_Resolution}.  Numerical truncation error is
seen to be irrelevant, except perhaps at the lowest resolution.  This
result agrees with the results of~\cite{Santamaria:2010yb} for the
{\tt Llama} code, and~\cite{Hannam:2010} for the {\tt BAM}
code, although~\cite{Hannam:2010} considers only the numerical waveform and not a hybrid. 

Turning to the gravitational wave extraction at finite radius $R_{\rm
  GW}$, we construct hybrid waveforms based on NR waveforms at two
finite extraction radii, and one where the waveform is extrapolated to
infinite extraction radius (we use a TaylorT3 PN waveform with
$M\omega_m=0.042$).  The solid lines in
Fig.~\ref{fig:Numerical_Resolution} report the differences between the
``finite-radius hybrids'' and the hybrid based on the extrapolated NR
waveform.  The effects of finite-radius wave-extraction are
significant, making even the $R_{\rm GW}=385M$ numerical simulation
unusable for parameter estimation at moderately large $\rho_{\rm
  eff}$.  Thus, it is imperative to always utilize waveforms
extrapolated to infinite extraction radius, or to use Cauchy
characteristic extraction \cite{Reisswig2009}.  This result is of
relevance to other studies such as~\cite{Santamaria:2010yb}, because
most of the NR waveforms used in that analysis have a finite
extraction radius. One should note that the results presented here
are specific to the numerical waveforms we study, and may not hold in
general.  For example, the results from~\cite{HannamEtAl:2010} show a 
lower error for finite extraction radius waveforms, but were obtained with a different evolution system and gauge (moving puncture BSSN, rather than Generalized Harmonic).

Wave extrapolation to infinite extraction radius is already well
established~\cite{Boyle2007,Boyle-Mroue:2008}, so one might view an
investigation of the impact of finite $R_{\rm GW}$ as less important.
However, we can use the finite $R_{\rm GW}$ waveforms to comment on
several topics of relevance.

\begin{figure}
\centerline{\includegraphics[scale=0.425]{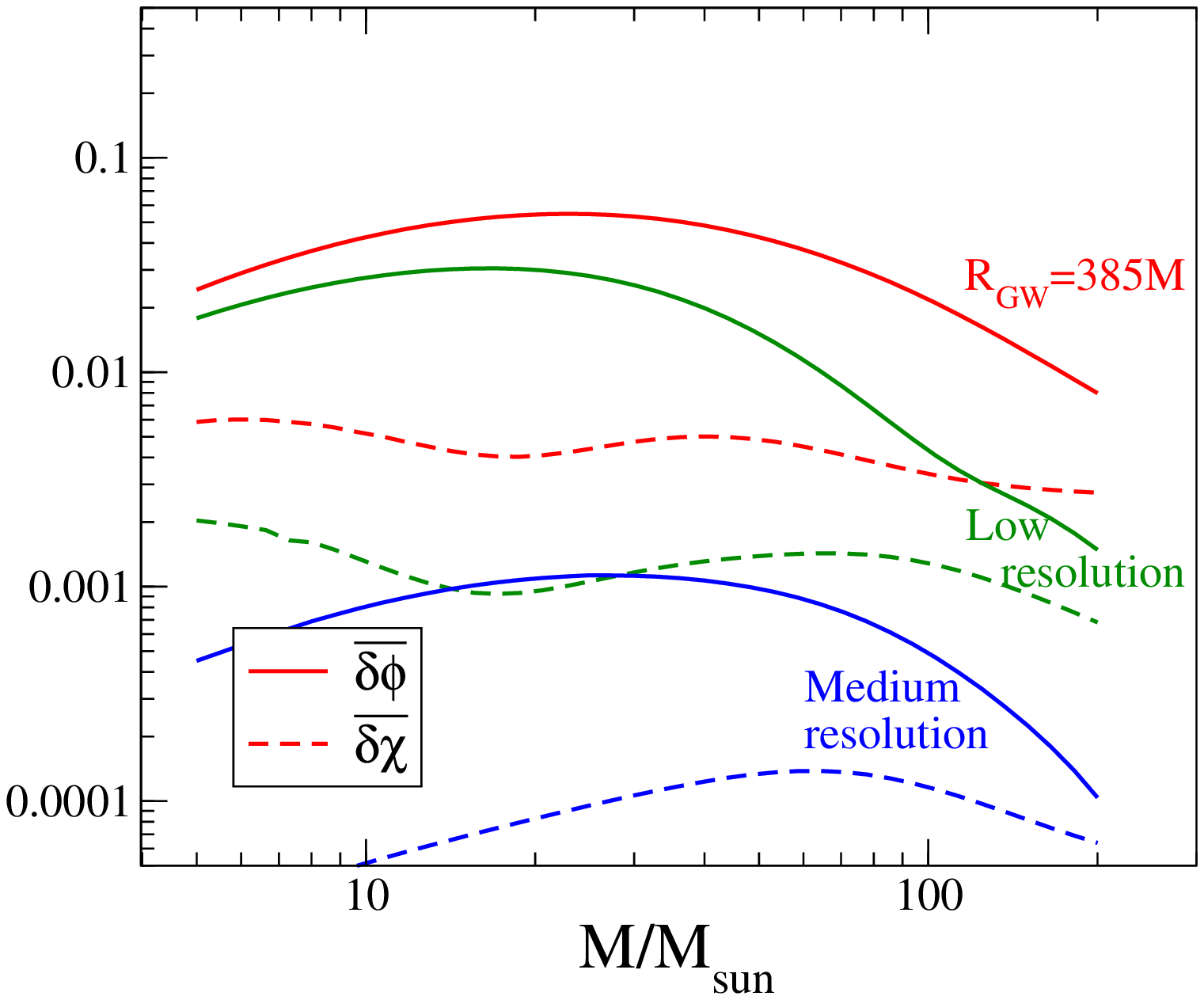}
\includegraphics[scale=0.425]{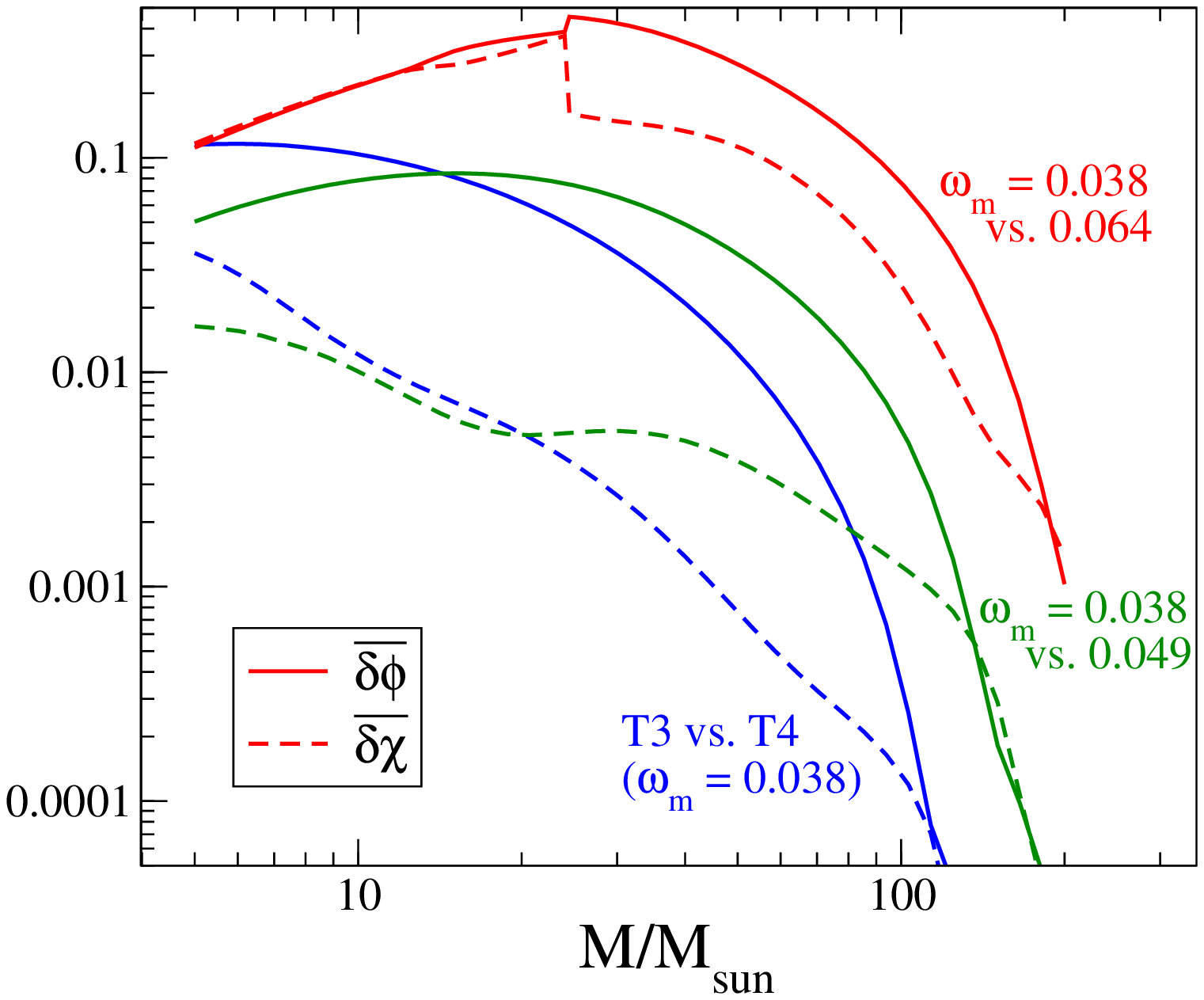}}
\caption{\label{fig:Lindblom_AmpPhaseErr} Contribution of amplitude--
  and phase--errors  to the overall error $\Q$.  {\bf Left:}
  the effect of wave-extraction at finite radius $R_{\rm GW}=385M$,
  and the effect of numerical truncation error (see Fig.~\ref{fig:Numerical_Resolution}). {\bf Right:} the effect of a change of
  Post-Newtonian approximant (TaylorT3 vs. TaylorT4), and changes to the
  matching region ($\omega_m = 0.038$ vs 0.049 and 0.064), cf. Sec.~\ref{sec:44}.}
\end{figure}

 We begin by investigating the relative importance of phase--
  and amplitude--errors.  Using the Fourier-transforms
  $\widetilde{\delta h}(f)$ and $\tilde{h}(f)$, we follow Lindblom et
  al.~\cite{Lindblom2008} and write $\widetilde{\delta
    h}(f)=\big[\delta\chi(f)+i\delta\phi(f)\big]\tilde h(f)$, so that
  $\delta\chi(f)$ and $\delta\phi(f)$ represent the fractional
  amplitude--error and the phase--error of the waveforms in the
  Fourier-domain.  Defining noise-weighted averages,
  $\overline{\delta\chi}=||\delta\chi\, h || / ||h||$,
  $\overline{\delta\phi}=||\delta\phi\, h || / ||h||$ (where
  $||\delta\chi\, h||$ is obtained by substituting $\delta\chi(f)\tilde
  h(f)$ into the integral in Eq.~\ref{eq:NWIP}), Lindblom
  et. al. show that
\begin{equation}\label{eq:Lindblom_AmpPhaseErr}
\frac{||\delta h||}{||h||} = \sqrt{\; \overline{\delta\phi}^2 + \overline{\delta\chi}^2\,}.
\end{equation}
Fig.~\ref{fig:Lindblom_AmpPhaseErr} plots $\overline{\delta\chi}$
and $\overline{\delta\phi}$ for three representative comparions
considered in this paper.  

In the case of numerical errors the phase errors dominate over the
  amplitude errors, as can be seen in the left panel of
  Fig.~\ref{fig:Lindblom_AmpPhaseErr}. The right panel of
  Fig.~\ref{fig:Lindblom_AmpPhaseErr} shows the amplitude and phase
  errors for a change in PN approximant and for different
  matching frequencies, which we will discuss in more detail in
  Sec.~\ref{sec:44}.  In these cases, amplitude errors sometimes are
  similar to phase errors, especially for very high matching
  frequencies or very large total mass. In general, however, it is fair to make the assumption that
  phase errors dominate over amplitude errors.  The discontinuity in $\overline{\delta\chi}$ and $\overline{\delta\phi}$ for one of the plotted camparisons arises because the best-fit alignment between the two hybrids being compared discontinuously jumps between two local minima as $M$ passes through a critical value where the global minimum jumps from one to the other minimum.

\begin{figure}
\centerline{\includegraphics[scale=0.4]{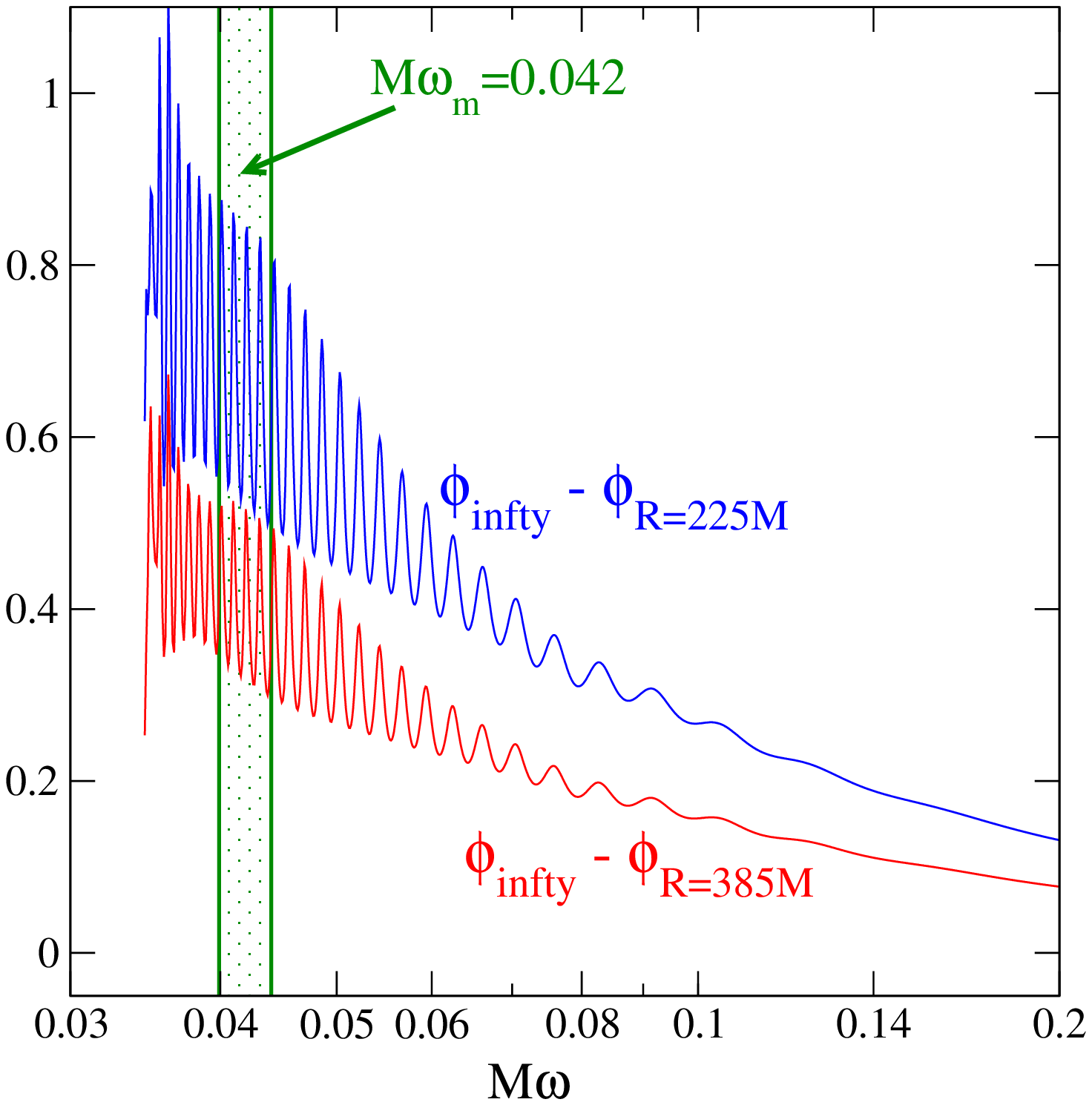}
\includegraphics[scale=0.4]{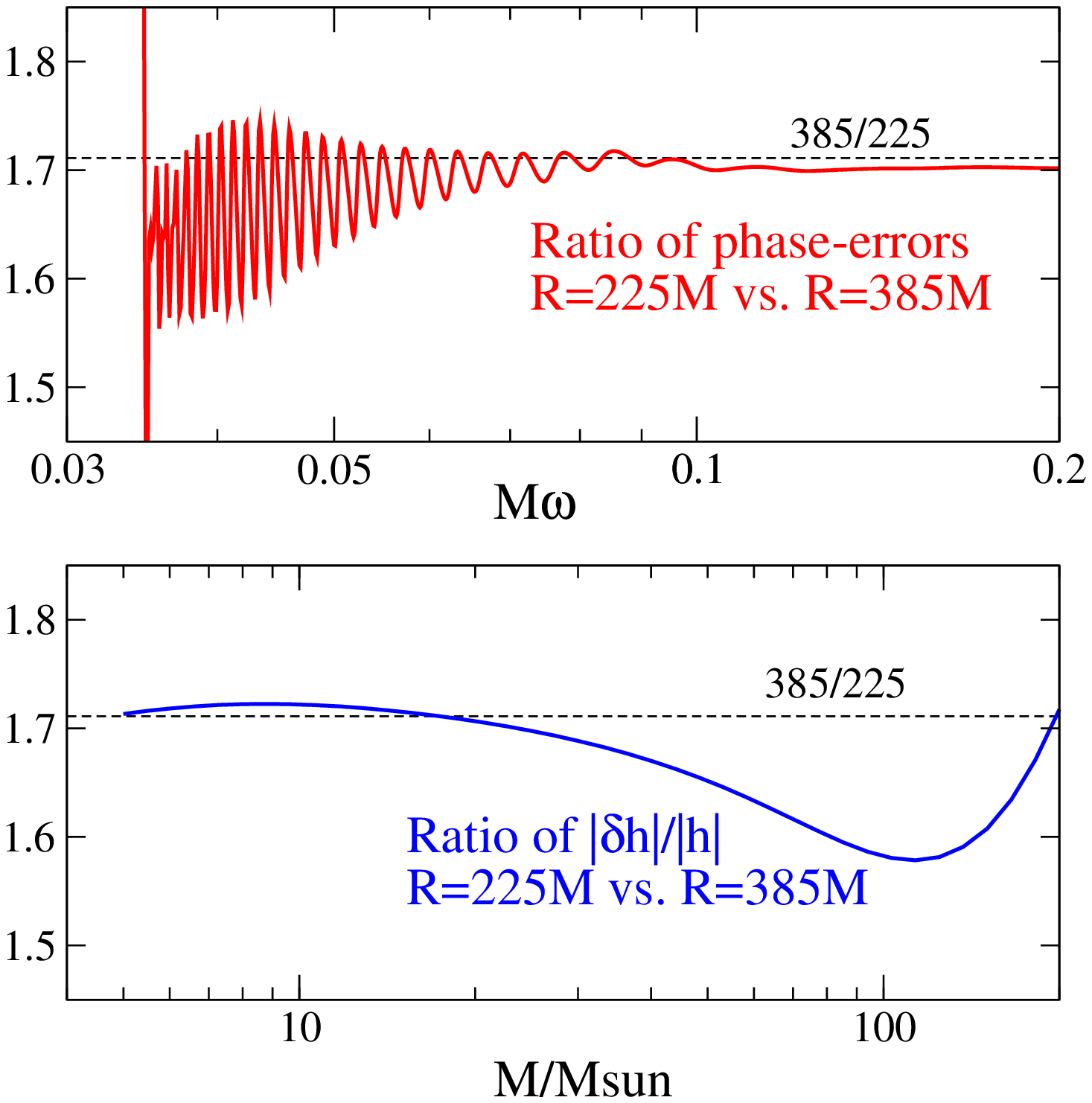}}
\caption{\label{fig:SystematicErrors} {\bf Left:} Phase difference
  between the NR waveforms extracted at finite $R_{\rm GW}$ relative
  to a waveform extrapolated to infinite extraction radius. {\bf
    Right:} ``Convergence test" of finite $R_{\rm GW}$ waveforms with
  extraction radius.  The top right panel shows the phase error, the
  bottom panel $\Q$, and both are seen to decay like $1/R_{\rm GW}$. }
\end{figure}

Next, we recall that numerical simulations are particularly
susceptible to phase errors during the inspiral phase, where the
energy flux is small and the inspiral time-scale is long.
Finite-radius gravitational waveforms induce errors with similar
properties, as can be seen in the left panel of
Fig.~\ref{fig:SystematicErrors}, which plots the phase-errors of the
finite $R_{\rm GW}$ numerical waveforms.  Therefore, finite-extraction
waveforms are a good model to assess the importance of phase-errors
during the inspiral phase.  Comparing the left panel of
Fig.~\ref{fig:SystematicErrors} to
Fig.~\ref{fig:Numerical_Resolution}, we find that phase-errors of
$\sim 0.5$ rad during the numerical simulation -- whatever their
origin --  are not indistinguishable in advanced detectors and should be avoided.

The right panels of Fig.~\ref{fig:SystematicErrors} demonstrate that
the phase-errors and the respective $\Q$ values are proportional to
$1/R_{\rm GW}$.  This confirms our earlier assertion that $\Q$ is
proportional to the dominant error that enters a hybrid
waveform (cf. the discussion after Eq.~(\ref{eq:dh-over-h-first})). This proportionality allows us to predict how small numerical
phase-errors should be, namely $\lesssim 0.2$ rad to be undetectable at
$\rho_{\rm eff}=40$, and proportionally smaller for higher SNR.  To include some
safety margin, we recommend a target of about $0.1$ rad phase error
when plotted similarly to Fig.~\ref{fig:SystematicErrors} as a
function of frequency.  The tolerable phase error during the
inspiral will naturally depend on many factors, most notably the length of the
NR waveform (i.e., the hybridization frequency $\omega_m$).  We have
not investigated scaling with length of the NR waveform, so our
recommendation only holds formally for $M\omega_m\approx 0.04$.

\begin{figure}
\centerline{\includegraphics[scale=0.36]{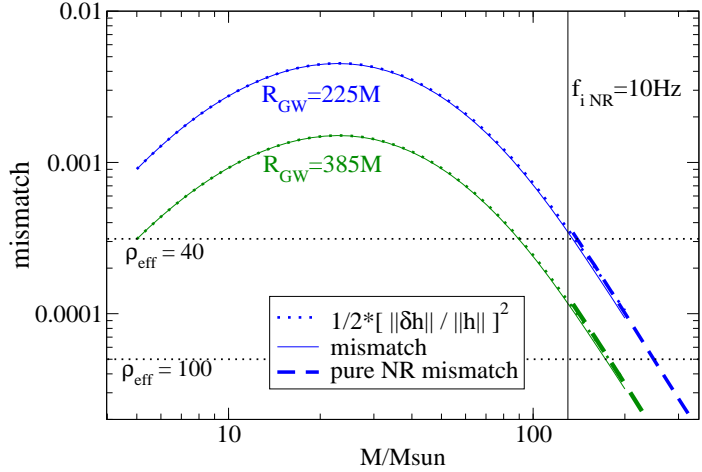}}
\caption{ \label{fig:mismatch_extracrad} Verification of
  Eq.~(\ref{eq:mismatch}) relating $\Q$ and mismatch.  For the finite
  radius hybrids shown in Figs.~\ref{fig:Numerical_Resolution}
  and~\ref{fig:SystematicErrors}, we plot mismatch ${\cal M}$ and
  $(\Q)^2/2$, which are virtually identical.  We also plot mismatches
  between pure NR waveforms (no hybrids) for so high masses that the
  initial GW frequency of the NR waveform is $f_{i \rm NR}\ge 10$Hz.}
\end{figure}

Fig.~\ref{fig:mismatch_extracrad} uses the comparison between finite
$R_{\rm GW}$ and extrapolated waveforms to motivate statements
that we made
in the context of Eq.~(\ref{eq:Q-missmatch-approx}).  We plot ${\cal
  M}$ and $\left( \Q \right)^2/2$.  These quantities should agree if $h_{\parallel}$
is negligible, which indeed is the case. 

Finally, Fig.~\ref{fig:mismatch_extracrad} also shows mismatches
between the pure NR waveforms (not hybridized) at finite and infinite
extraction radius.  These mismatches are only meaningful for
sufficiently high mass $M$~so that the NR waveform begins at
frequencies, $f_{i \rm NR}$, below the Advanced LIGO frequency band at $10$Hz, i.e. for
$M>130M_\odot$. As expected, we see that for $M>130M_\odot$ the pure NR mismatches
agree very well with the hybrid mismatches, because
hybridization is unnecessary and unimportant for such high masses.
However, at lower masses the hybrids have much larger errors than at
higher masses.  The mismatch, for instance, reaches a maximum 10 times larger than the maximum mismatch of the
NR-only waveforms.  Therefore, one can use pure NR mismatches only to
ascertain the usability of NR waveforms for high masses.  Small
mismatches of pure NR waveforms at high masses {\em do not imply}
small mismatches of hybrid waveforms at lower masses.  The only way to
assess the quality of NR waveforms that are intended for use in hybrids lies in the construction of hybrids and computing errors based on these hybrids.  Pure NR mismatches are commonly used because of
their convenience (e.g., in the Samurai project~\cite{Hannam:2009hh}),
but one must keep their limitations in mind. 

All hybrids used in this section use the TaylorT3 PN approximant.
We have repeated the calculations with other approximants without any
change in our results. Our findings are, therefore, entirely due to the
behavior of the numerical waveforms.

\subsection{Choice of matching frequency}
\label{sec:44}

This section addresses at what frequency hybridization should occur,
i.e.,  how long the NR simulations must be.  There is a trade-off
between computational expense and accuracy. On the one hand, we wish
to match PN and NR waveforms as close to merger as possible to reduce
the length and computational cost of NR simulations. On the other
hand, PN waveforms diminish in accuracy toward merger.  Because of its
importance, this is a very active research topic~\cite{Santamaria:2010yb,Hannam:2010,Damour:2010}.

\begin{figure}
\includegraphics[scale=0.4]{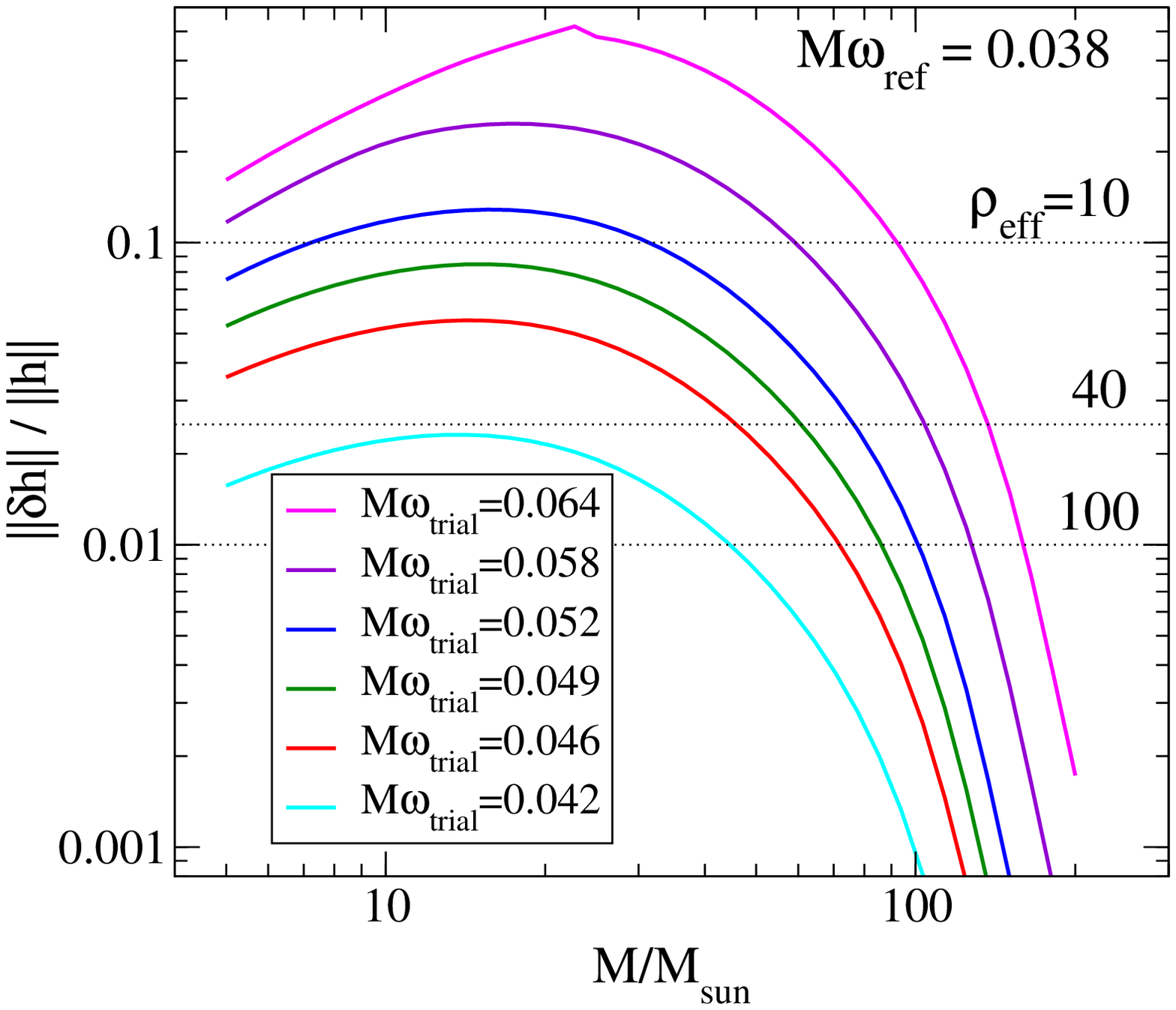}
$\;\;\;\;$
\includegraphics[scale=0.4]{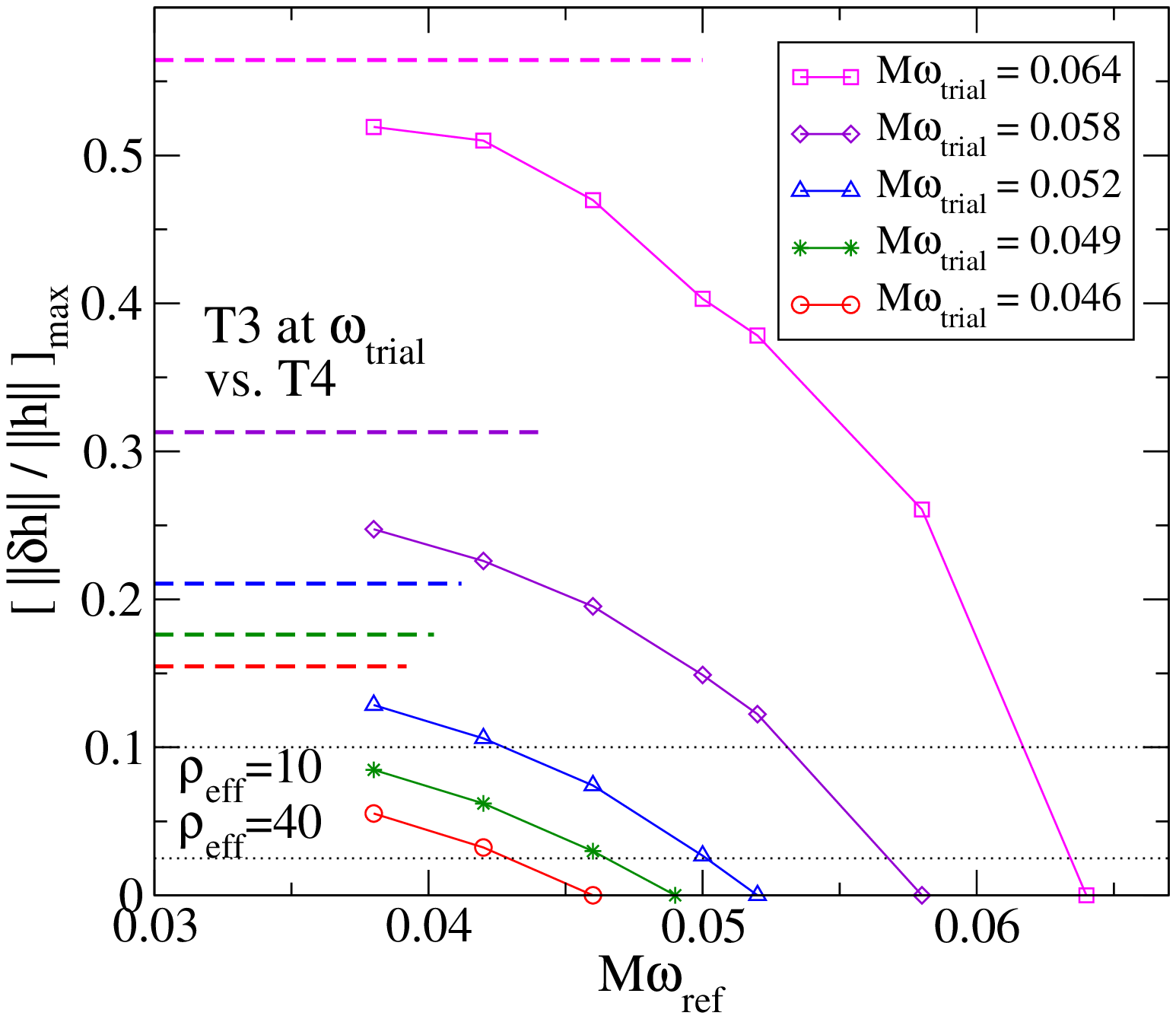}
\caption{\label{fig:DifferentMatchingFrequencies} Errors of TaylorT3
  hybrids, as a function of matching frequency $\omega_m$. {\bf Left:}
  $\Q$ between matching frequencies $M\omega_{\rm trial}$ and fixed
  $M\omega_{\rm ref}=0.038$.  The dotted lines indicate upper limit of
  $\Q$ for two different effective signal to noise ratios $\rho_{\rm
    eff}$.  {\bf Right:} $\Q$ maximized over $M$, as a function of the
  matching frequency of the reference hybrid for trial hybrids using
  several values of $\omega_{\rm trial}$.  The dashed lines are the
  maximum values of $\Q$ for each trial hybrid compared to a
  TaylorT4-hybrid matched at $M\omega_m = 0.042$. }
\end{figure}

As a first step towards understanding the importance of the matching
frequency $\omega_m$, we consider a series of hybrids where TaylorT3
is matched to the NR simulation at different values $\omega_m$, using
a matching interval $\delta\omega=0.1\omega_m$.  We then compute the
error measure $\Q$ between a ``reference hybrid'' which is matched at
the lowest possible frequency ($M\omega_m=0.038$), and ``trial
hybrids'', matched at higher frequencies.  The results are presented
in the left panel of
Fig.~\ref{fig:DifferentMatchingFrequencies}.  With increasing $\omega_m$ of
the trial hybrid, $\Q$ becomes unacceptably large.  With decreasing
$\omega_m$, $\Q$ decreases, as one would expect, and for $M\omega_{\rm
  trial}=0.042$, the error criterion appears to be satisfied even at $\rho_{\rm eff}=40$.  However, this decrease in $\Q$ is a combination of two
different origins. First, with earlier matching, the trial hybrid will
be closer to the exact waveform,  obtainable if one could
match arbitrarily early, $\omega_m\to 0$.  This is indeed the effect
that we are
trying to measure.  Second, as the matching frequencies of trial- and
reference-hybrid approach each other, the hybrid-waveforms themselves also approach each other, and $\Q$ will decrease.  To allow
conclusions about the trial hybrid, we must quantify the importance of
this artificial suppression of $\Q$.

The right panel of Fig.~\ref{fig:DifferentMatchingFrequencies}
attempts to determine the importance of the artificial suppression due
to close matching frequencies.  We fix $\omega_{\rm trial}$, and
consider decreasing values of $\omega_{\rm ref}$.  In the limit
$\omega_{\rm ref}\to 0$, the difference $\Q$ will measure the error in
the trial hybrid.  To reduce the dimensionality of the plot, the right
panel of Fig.~\ref{fig:DifferentMatchingFrequencies} shows the maximum
of $\Q$ over the mass-range $[5M_\odot, 200M_\odot]$.  As expected,
$\Q$ becomes smaller and approaches $0$ as $\omega_{\rm ref}\to\omega_{\rm
  trial}$. As was just alluded
to, this is the artificial suppression of $\Q$.  Let us now consider the opposite limit of $\omega_{\rm ref}\to
0$.  For large $\omega_{\rm trial}$ (for instance the top-most, magenta
curve ($M\omega_{\rm trial}=0.064$)), $\Q$ indeed levels off as
$\omega_{\rm ref}$ is made as small as possible, and asymptotes
towards what appears to be some maximum value. We interpret this
maximum value to represent the true error in the trial hybrid waveform
if our NR waveform were arbitrarily long.  For $M\omega_{\rm
  trial}=0.064$, this asymptotic value is $\Q\approx 0.5$, demonstrating
that a TaylorT3 hyrid matched at $M\omega_m=M\omega_{\rm trial}=0.064$
is not useful for any data-analysis purposes.   

As we move toward smaller $\omega_{\rm trial}$, the error $\Q$ falls.
Unfortunately, once $\omega_{\rm trial}$ is so small that the error
$\Q$ begins to get interesting (e.g. the lowest two curves,
$M\omega_{\rm trial}=0.046, 0.049$), the asymptotic behavior as
$\omega_{\rm ref}\to 0$ is no longer visible. In order to re-establish this behavior,
we would require such a small value in $\omega_{\rm ref}$ such that it is below the starting frequency of the numerical waveform. By utilizing only a {\em single} PN approximant, we can, therefore, learn only the following: the 15-orbit
SpEC waveform is long enough to ascertain that matching at high
frequencies $M\omega_m \gtrsim 0.055$ results in hybrids with
unacceptably large errors. However, the NR waveform is too short to
establish reliably the error incurred by matching at lower frequencies
$0.038\lesssim M\omega_m\lesssim 0.055$.

\begin{figure}
\includegraphics[scale=0.4]{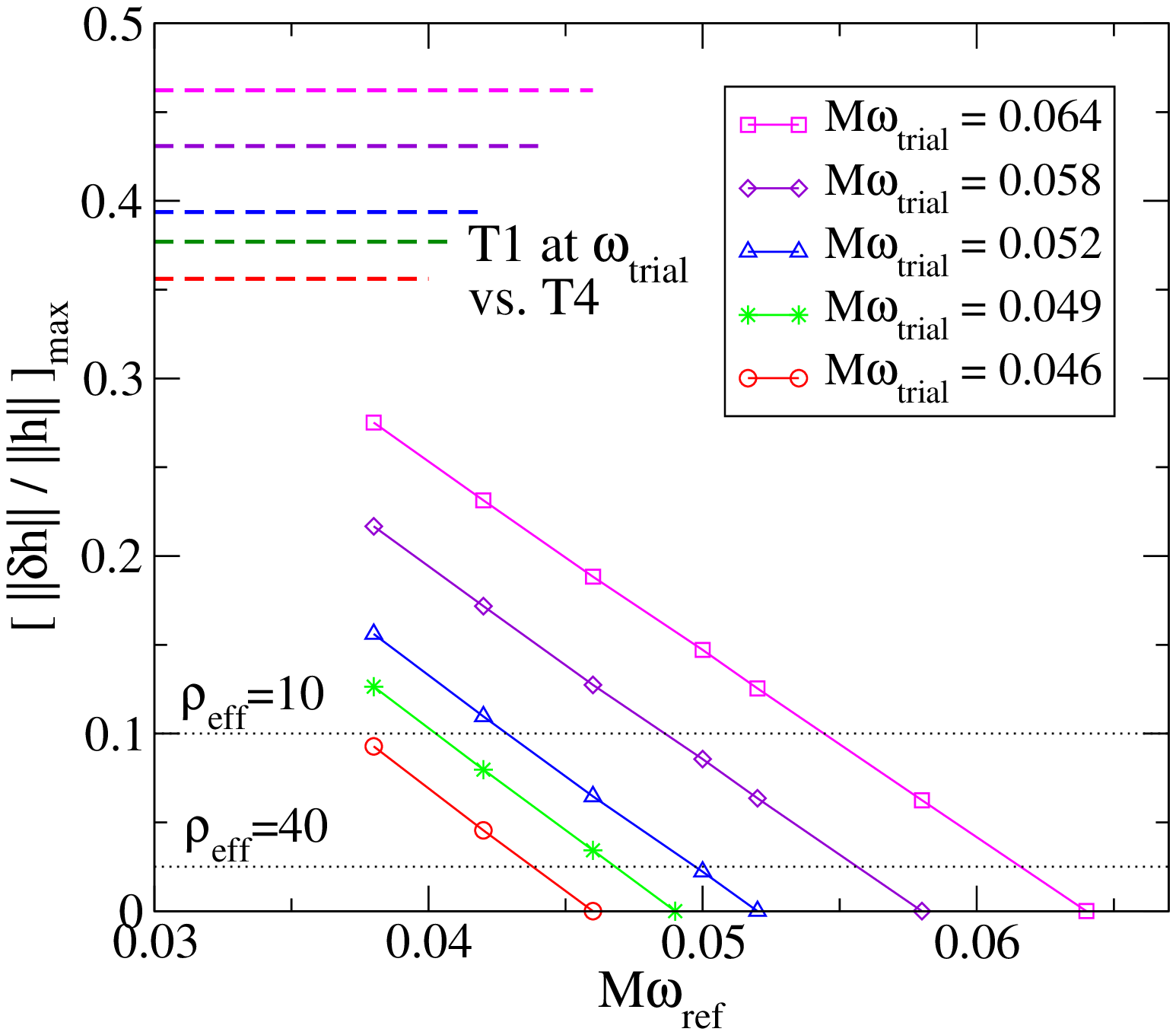}
$\;\;\;\;$
\includegraphics[scale=0.4]{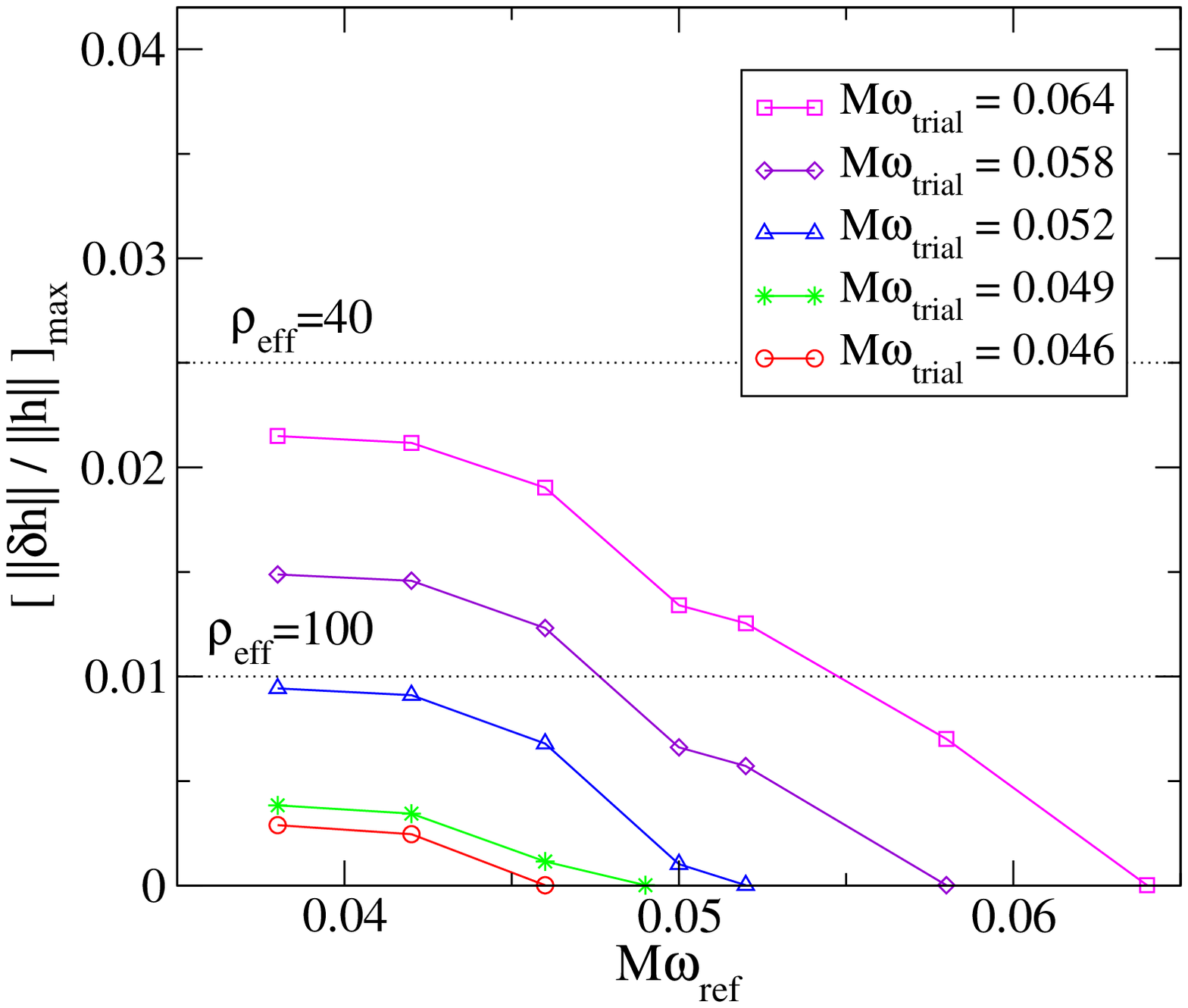}
\caption{\label{fig:DifferentPNmatching} {\bf Left:} Errors of {\bf
    TaylorT1} hybrids as a function of matching frequency.  {\bf
    Right:} Errors of {\bf TaylorT4} hybrids as a function of matching
  frequency.  These plots mirror the right panel of
  Fig.~\ref{fig:DifferentMatchingFrequencies}, except using a
  different PN approximant (note the different y-axis scale in the right panel).}
\end{figure}

To gain more insight into the effects shown in
Fig.~\ref{fig:DifferentMatchingFrequencies}, we explore hybrid
waveforms using different PN Taylor approximants.  A natural starting
point involves constructing our {\sl reference} hybrid using a PN
TaylorT4 approximant, which is known to be considerably more accurate
than TaylorT3 in the non-spinning equal mass case (see
Ref.~\cite{Boyle2007} for detailed discussion).  If we assume that
TaylorT4 incurs a negligible error relative to the ``true'' inspiral
waveform, error criteria $\Q$ between TaylorT4 hybrids and the
previously considered TaylorT3 hybrids will display the error in the
TaylorT3 hybrid.  These errors are indicated by the horizontal dashed
lines in the right panel of
Fig.~\ref{fig:DifferentMatchingFrequencies}.  (Because TaylorT4 agrees
so well with NR, for convenience here we use only one TaylorT4 hybrid matched
at $M\omega = 0.042$).  These dashed horizontal lines in the right panel
of Fig.~\ref{fig:DifferentMatchingFrequencies} are consistent with our
conclusions of the preceeding paragraph: the errors in the TaylorT3
hybrids matched at high frequency ($M\omega_{\rm trial}=0.058, 0.064$)
can be resolved by pushing $\omega_{\rm ref}$ to the start of the NR
waveform, and the asymptotic value of the TaylorT3-only comparisons
agrees with those of TaylorT3---TaylorT4. For smaller $\omega_{\rm
  trial}$, we cannot reach the asymptotic regime, and while $\Q$ for
TaylorT3-only comparisons increases as $\omega_{\rm ref}$ is pushed as
low as possible, the error is still far away from the expected limit
given by the TaylorT3-TaylorT4 comparison.  The dashed
TaylorT3---TaylorT4 lines indicate errors too large even for
$\rho_{\rm eff}=10$.  Under our conservative error
  criteria, we conclude that the NR
waveform is too short to construct useful TaylorT3 hybrids.

So far, we have used the TaylorT3 PN waveform, because its analytical
nature makes it the simplest to implement.  We now consider different PN approximants. The left panel of
Fig.~\ref{fig:DifferentPNmatching} plots the same information as the
right panel of Fig.~\ref{fig:DifferentMatchingFrequencies}, except
that TaylorT3 is now replaced by TaylorT1.  Similar to the right panel of
Fig.~\ref{fig:DifferentMatchingFrequencies}, we can clearly see the
suppression of the differences when $\omega_{\rm trial}$ is too close to
$\omega_{\rm ref}$.  As above, this effect manifests itself in
$\Q\propto |\omega_{\rm ref}-\omega_{\rm trial}|$.  However, as
opposed to TaylorT3, for the TaylorT1 approximants, we do not see any
indication that $\Q$ levels off as $\omega_{\rm ref}$ is made smaller,
not even for the largest possible difference $(\omega_{\rm trial}-\omega_{\rm ref})$.  Our conclusions for
TaylorT1 hybrids, therefore, are identical to those reached for TaylorT3
hybrids; the current 15-orbit NR waveform is too short to construct
reliable TaylorT1 hybrids on the basis of our
  error criteria. Or conversely, TaylorT1 is too inaccurate
to be utilized with the currently available numerical
waveforms within our error framework. 

The right panel of Fig.~\ref{fig:DifferentPNmatching} shows the same
comparisons using the TaylorT4 PN approximant.  TaylorT4 is special,
in that it happens to agree very well with the numerical simulations
for the equal-mass, non-spinning case.  Indeed, the right panel of
Fig.~\ref{fig:DifferentPNmatching} confirms this fact; the errors $\Q$
are far smaller, within our limits for detection 
and parameter
estimation even for $\rho_{\rm eff}=100$ for matching frequencies $M\omega_m < 0.052$.  For all
$\omega_{\rm trial}$, convergence to an asymptotic value is visible as
$\omega_{\rm ref}$ is decreased.  Therefore, TaylorT4 hybrids
are suitable in the equal-mass, non-spinning
case for event detection and parameter
estimation.  However, as is well known, this agreement is coincidental
and does not carry over to more generic
configurations~\cite{Hannam2007c}; the exceptional good behavior of
TaylorT4 is, thus, not helpful for the generic case. Nevertheless, convergence towards
an upper error limit at lower matching frequency suggests the suitability of different PN approximants at frequencies lower
than current NR waveforms can provide.

\begin{figure}
\centerline{\includegraphics[scale=0.5]{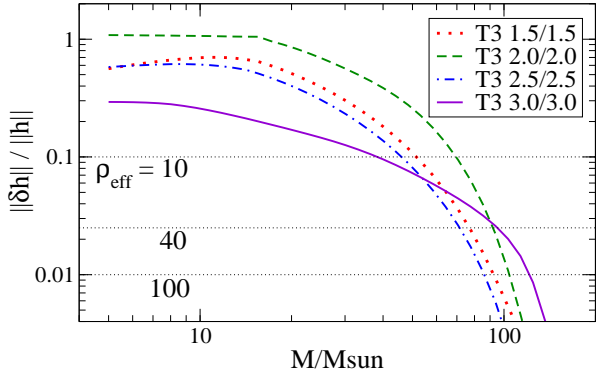}}
\caption{\label{fig:DiffPNs} Error criterion evaluated between
  TaylorT3 hybrids at different post-Newtonian order.  All hybrids
  matched at $M\omega_m = 0.042$, and all comparison relative to a
  hybrid using Taylor T3 3.5/3.0 (3.5PN order in phase, 3.0 PN order
  in amplitude).}
\end{figure}

We point out that our results presented in
Figs.~\ref{fig:DifferentMatchingFrequencies}
and~\ref{fig:DifferentPNmatching} are consistent with the results of
Buonnano {\sl et al.}~\cite{Buonanno:2009} who compute
overlaps for $M = 20 M_\odot$ between non-hybridized Taylor
approximants terminated at ISCO.  Translating into our notation, they
find $\Q = 0.64$ when comparing Taylor T4 to Taylor T3, and $\Q =
0.34$ when comparing Taylor T4 to Taylor T1.  At our highest matching
frequency $M\omega_{\rm trial}=0.064$ (which is closest to pure PN
waveforms), one can read off from Figs.~\ref{fig:DifferentMatchingFrequencies}
and~\ref{fig:DifferentPNmatching} that $\Q=0.56$ and $\Q=0.46$
respectively. This agreement is reasonably good, considering the
differences in the examined waveforms.

To close this section, we briefly investigate hybrids constructed from
TaylorT3 at different PN orders.  We construct all hybrids
at a matching frequency of $M \omega_m = 0.042$ and compute $\Q$
between a reference hybrid using the most accurate PN order (3.5PN in
phase, 3PN in amplitude) and trial hybrids with lower PN
expansion orders.  The legend of Fig.~\ref{fig:DiffPNs} indicates which
PN orders we consider.  All differences, even between 3.0 and 3.5PN
phase accuracy, are so large, suggesting that lower order PN
approximants may not be sufficiently accurate for data-analysis purposes.  This finding
is consistent with our earlier findings; we established earlier that
using our error criteria, TaylorT3 3.5/3.0 is insufficient when matched at
$M\omega_m=0.042$.  As Fig.~\ref{fig:DiffPNs} confirms, lower-order PN approximants are even less
accurate, and consequently, are also insufficient within our analysis.

\subsection{ Effects of the noise curve}
\label{sec:noise-curve}

We have so far considered detection and measurement accuracies only in
the context of a single Advanced LIGO interferometer. This section
briefly presents $\Q$ computations for different instrument noise
curves including Initial LIGO (with a low-frequency cut-off of 40Hz) and a frequently used analytical fit~\cite{Ajith:2008b} to
the Advanced LIGO sensitivity curves.  These noise curves are plotted
in Fig.~\ref{fig:aligonoise}.  The overall scale of the noise curve
$S_n(f)$ and the overall scale of the waveforms $h(t)$ cancels in the
normalized error criterion $\Q$.  Nevertheless, the shape of the
waveforms and of $S_n(f)$  are important~\cite{Damour:2010}.  Specifically, a
noise-spectrum with a wider bandwidth will be sensitive to a larger
portion of the waveform.  

\begin{figure}
\centering
\includegraphics[scale=0.5]{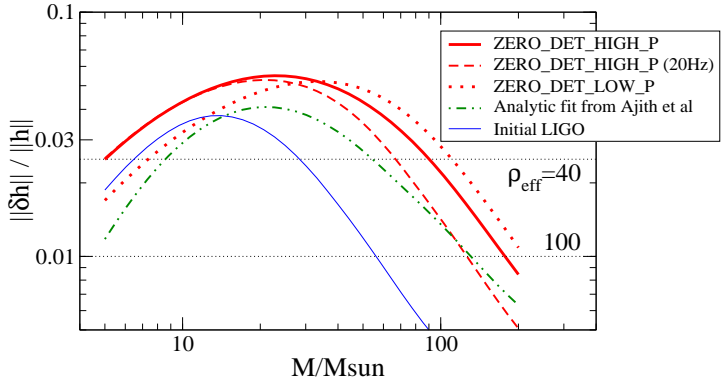}
\caption{\label{fig:MatchingAllNoise} Effect of different noise curves
  on $\Q$.  Shown are differences between finite extraction radius
  $R_{\rm GW}=385M$ and waveforms extrapolated to infinity,
  cf. Fig.~\ref{fig:Numerical_Resolution}.}
\end{figure}

In order to investigate the impact of different noise curves, we use the waveform at finite extraction
radius $R_{\rm GW}=385M$, already presented in
Figs.~\ref{fig:Numerical_Resolution}--\ref{fig:mismatch_extracrad}.  Fig.~\ref{fig:MatchingAllNoise} repeats this earlier calculation for all noise curves presented in Fig.~\ref{fig:aligonoise}.   The error $\Q$ is larger for the larger-bandwidth
noise curves, in particular {\tt ZERO\_DET\_HIGH\_P} and {\tt
  ZERO\_DET\_LOW\_P}.  Comparing {\tt ZERO\_DET\_LOW\_P} and {\tt ZERO\_DET\_HIGH\_P}, we note that {\tt ZERO\_DET\_LOW\_P} is more
sensitive at low frequencies relative to high frequencies, and
correspondingly it results in a larger $\Q$ at high masses (i.e., low-frequency waveforms) and smaller $\Q$ at low masses, compared to {\tt
  ZERO\_DET\_HIGH\_P}.  We also include the {\tt ZERO\_DET\_HIGH\_P}
noise curve with a low-frequency cutoff at $20$ Hz.  This increased
low-frequency cutoff has no impact on low mass systems with $M\lesssim
20M_\odot$, but results in reduced sensitivity (i.e. smaller $\Q$) at
larger masses, compared to the unmodified {\tt ZERO\_DET\_HIGH\_P}
noise curve.

We note that the analytical fit to the Advanced LIGO noise curve
(green line) underestimates the error $\Q$ compared to the projected
Advanced LIGO noise curve. Also, hybrid waveforms, sufficient for
measurement accuracy with Initial LIGO, may no longer meet the
  same measurement
criteria within an identical mass range and $\rho_{\rm eff}$ in the case
of Advanced LIGO, as seen in Ref.~\cite{Damour:2010}.
Besides the larger bandwidth (which directly affects $\Q$), the
Advanced LIGO noise curves are more sensitive than Initial LIGO.  This
needs to be taken into account when choosing an acceptable $\rho_{\rm
  eff}$. The effect of noise curve has also been studied in Ref.~\cite{Santamaria:2010yb}, 
where the error between different resolutions of NR waveforms and different
types of NR codes is calculated through the Advanced LIGO and Initial LIGO noise curves. A similar trend is observed with respect to an increase in sensitivity (and therefore error) at high masses.

\section{Discussion}
 \label{sec:Discussion}

 In this paper we have presented a comprehensive study of errors that
 affect hybrid waveforms.  We assess the quality of the hybrids 
 based on their suitability for Advanced LIGO parameter estimation, using
 the criterion most recently presented in~\cite{Lindblom2008}, 
and refined in~\cite{Damour:2010},
\begin{equation}\label{eq:Q-discussion}
\frac{\norm{\delta h}}{\norm{h}} < \frac{1}{\rho_{\rm eff}}.
\end{equation}
The left-hand side of this inequality is independent of the source
distance, and therefore, forms a convenient quantity for this
analysis. The right-hand side incorporates the single detector signal-to-noise ratio
$\rho$ of to-be-analyzed events, a safety factor $\varepsilon$~\cite{Damour:2010}  and possibly a correction factor to account for a
network of detectors, cf. Eq.~(\ref{eq:rhoeff}). The inequality~(\ref{eq:Q-discussion}) provides a bound on $\rho$ (or,
equivalently, distance) below which the error $\delta h$ is
undetectable in the GW detectors, indication that $h$
is sufficiently accurate.  As argued in
Sec.~\ref{sec:QuantifyingErrors}, $\rho_{\rm eff}=40$ is a reasonable
value, on which the following conclusions are based. However, one
should keep in mind that $\rho_{\rm eff}=100 - 250$ might be relevant
if optimistic assumptions about the number of GW sources turn out to
be correct, resulting in stronger accuracy demands on hybrid waveforms.

We investigated a large variety of effects that cause
possible errors
$\delta h$.  The general theme of our findings is that
Eq.~(\ref{eq:Q-discussion}) places strong
accuracy requirements on all elements of the hybrid construction:
\begin{enumerate}
\item
Section~\ref{sec:41} shows that the PN waveform has to be aligned
correctly to $\delta t_c\lesssim 1M$ time-offset, about 1/100 of a gravitational
wave cycle.
\item Section~\ref{sec:43} shows that the numerical waveform
must have phase-errors $\lesssim 0.1$ radians. 
\item  Section~\ref{sec:44}
indicates that PN Taylor--approximants
in the time domain are not sufficiently
  accurate when representing the inspiral waveform up to the start of currently
available numerical relativity waveforms.     
\end{enumerate}

These three findings are interrelated.  We have already pointed out in
Eq.~(\ref{eq:t_35PN}) that the 3.5PN order contributes
a time-offset of about $t_{3.5\rm PN}\approx 100M$ at typical matching
frequencies $M\omega\approx 0.04$.  The analytical formulae from the
TaylorT2 approximant~\cite{Damour:2000zb,Damour:2002kr} show that the 3.5PN order
for an equal-mass binary contributes
\begin{equation}
\phi_{3.5\rm PN}=-\frac{1}{4}\frac{357,185}{7,938}\pi (M\Omega)^{2/3}
\approx - 2.6 \, \, \mbox{rad}
\end{equation}
to the GW phase (at orbital frequency
$M\Omega=M\omega/2=0.02$).  Both the PN contributions $t_{3.5\rm PN}$ and $\phi_{3.5\rm
  PN}$ are much larger than the limits on the time and phase errors we
established for the numerical waveforms ($1M$ and $0.1$ rad,
respectively), so it is unsurprising
that the time-domain Taylor approximants are insufficient, or conversely, that the
numerical waveforms are too short within our strict error requirements.

\begin{figure}
\centerline{
\includegraphics[scale=0.38]{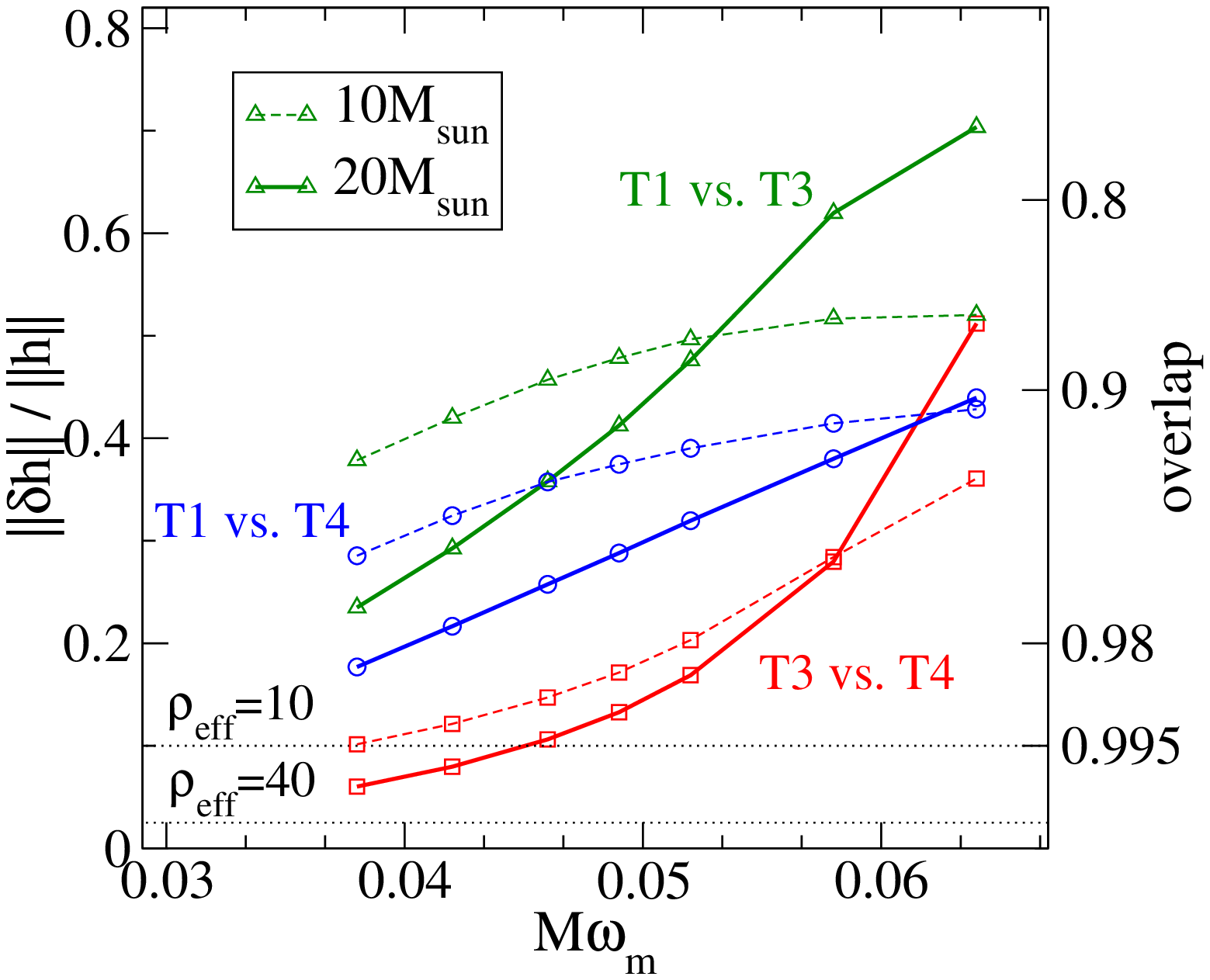}
$\;\;\;\;$
\includegraphics[scale=0.38]{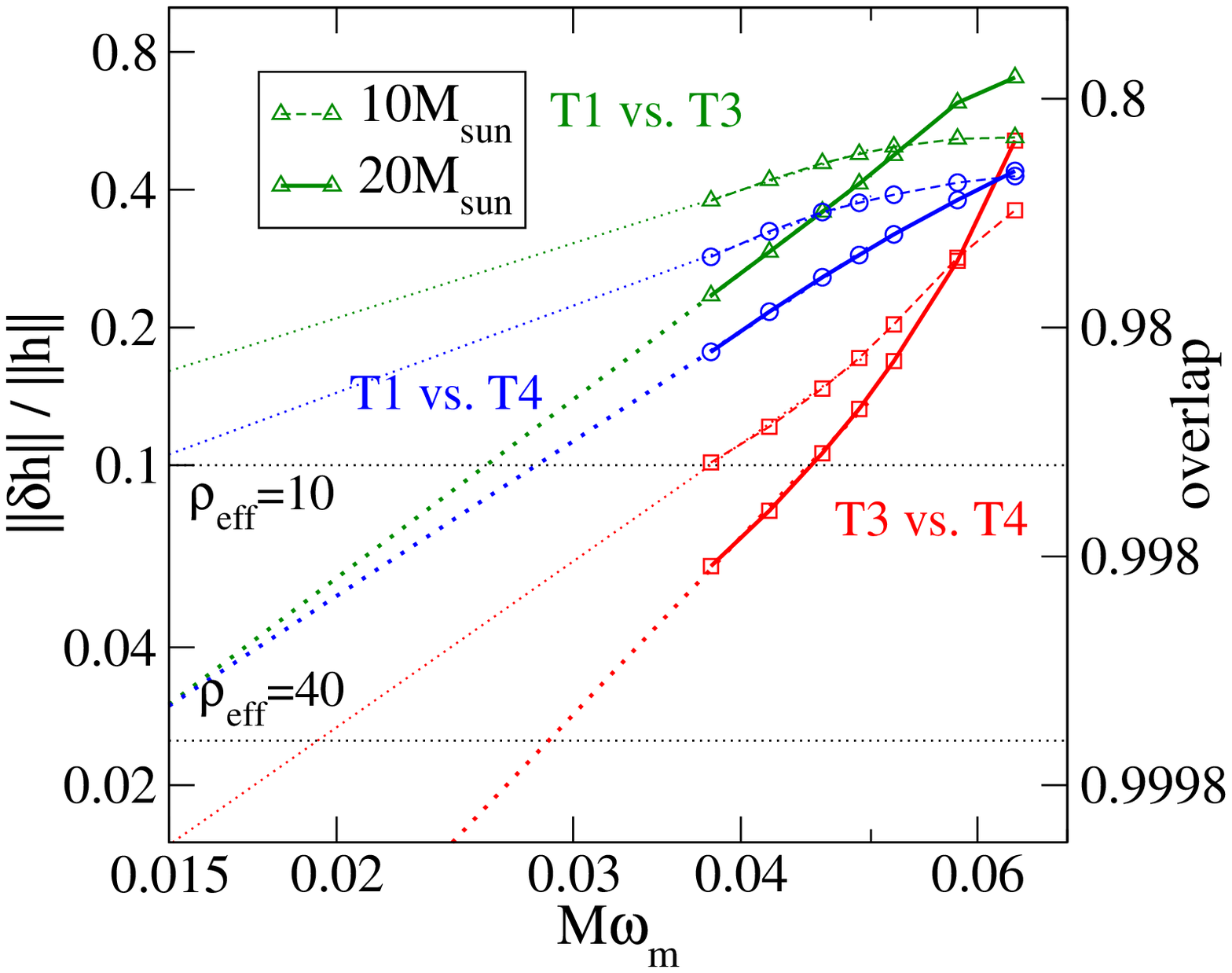}}
\caption{\label{fig:T1vsT3vsT4_matching}
  Differences between hybrids using different PN approximants, but
  matched at the same matching frequency.  Plotted is the
  value of $\Q$ at $10M_\odot$ and $20M_\odot$, as a
  function of $\omega_m$.  Both plots show the same data, albeit with
  different axes.  The right plot contains power law fits to the
  smallest four data-points.}
\end{figure}

So how long do NR waveforms need to be in order to reliably attach a
Taylor-PN inspiral?  Or equivalently, how early must one stop using
the PN waveforms?  Fig.~\ref{fig:T1vsT3vsT4_matching} suggests a 
conservative answer to this question.  Both panels of
Fig.~\ref{fig:T1vsT3vsT4_matching} show the same data, but with
different axes.  The data presented are $\Q$ for $M=10M_\odot$ and $M=20M_\odot$ between pairs of
\{TaylorT1, TaylorT3, TaylorT4\} hybrids, all matched at the same
frequency $\omega_m$, and plotted as a function of hybridization frequency $\omega_m$.  Because the TaylorT hybrids only differ in
their treatment of uncontrolled higher-order PN terms, these
differences are a reasonable measure of the truncation error of the PN
series.  Fig.~\ref{fig:T1vsT3vsT4_matching} shows
that these differences are very large, far above the limit of
Eq.~(\ref{eq:Q-discussion}).  The right panel of
Fig.~\ref{fig:T1vsT3vsT4_matching} attempts power-law fits to $\Q$ as
a function of $\omega_m$.  Extrapolation based on fits is highly
unreliable, but nevertheless, this figure indicates that one must match at
significantly lower frequencies $\omega_m$ than currently available.
The time-to-merger scales in proportion to $(M\omega)^{-8/3}$, so this
implies a need for NR waveforms several times
longer than those currently computed.

Figure~\ref{fig:T1vsT3vsT4_matching} shows data for total mass
  of $10M_\odot$ and $20M_\odot$, with the lower total mass resulting
  in larger $\Q$.  Our findings are compatible with the results of
earlier PN-inspiral and EOB work~\cite{Buonanno:2009}. In the case
of the significantly lower mass (1.42, 1.38) $M_{\odot}$, using their
faithfulness quantity, Buonnano {\sl et al.} show that Taylor T2, T3,
and T4 exhibit a corresponding $\Q$ value of 0.14 with each other,
whereas Taylor T1 has a somewhat worse agreement of 0.24.  However, we
reiterate that the criterion~(\ref{eq:Q-discussion}) is sufficient,
and may therefore place unnecessary strong constraints on
waveform accuracy.  

Figure~\ref{fig:T1vsT3vsT4_matching} clearly
presents a challenge for the current state of the art in numerical and analytical waveform modelling.  Fortunately, there
are several potential avenues to reduce the ``frequency gap''  (in the language
of~\cite{Damour:2010}) between PN and NR waverforms.

First, one can compute longer NR waveforms. This is
  clearly challenging, as the following considerations based
    on the {\tt SpEC} code illustrate.   By order of magnitude,
  a simulation of a generic inspiral with moderate mass-ratio and
  spins currently requires several $10^4$ CPU-hours, lasting
   several months.  These numbers apply to an
  evolution time of about $5000M$ at accuracies indicated by our
  analysis here.  Running {\tt SpEC} for $\sim 30$
    orbits is feasible, however, this will require wall-clock times approaching
    a year.  We caution that no careful convergence tests have been
    performed for such simulations, though preliminary simulations
    appear promising.  Excessive wall-clock time is the most
    restrictive factor in simulations longer than $\sim 15$ orbits.
    This situation can be ameliorated with at least
  three different approaches: increase parallelism to many more cores;
  utilize a faster computing architecture, e.g. graphics
  accelerators~\cite{MroueGpuTalk2010}; or develop more efficient
  numerical
  algorithms~\cite{LauPfeiffer2008,Hennig:2008af,LauLovelacePfeiffer2011}.
  All three approaches are promising, and we expect that a
  combination of them will result in a significant extension of
  computational capabilities.

An alternative approach is to improve the accuracy of the PN series, either through calculation of higher PN order, or by seeking more rapidly converging analytical representations of the
inspiral phase.  The EOB formalism provides such a
representation, as recently argued in~\cite{Damour:2010}.  
However, 
some improvements to the EOB formalism (e.g.,~\cite{DIN})
were developed after high-accuracy NR waveforms became
available for comparisons.  While the improvements are well motivated
on physical grounds, the additional confidence that EOB models are
developed independently from the numerical results no longer applies.
Finally, once one considers tuned models, one loses all ability to
use differences (e.g., to the numerical result) as an error measure.
Such models, by definition, will agree very well in the frequency
range in which they are tuned to numerical simulations.  However, such
tuning does not guarantee that the cycles {\em before} the tuning
region are well represented by the tuned model.  Even with these
reservations, EOB models are perhaps the most promising avenue to
 close the ``frequency gap'' between numerical simulations and Taylor
post-Newtonian models.

Finally, a more detailed analysis of waveform accuracies for
  GW data-analysis might result in weakened requirements.  The impact of
  detector calibration errors or of non-Gaussian detector-noise might
  overwhelm the limits presented here that are based on the ideal
  detector case.  Furthermore, parameter estimation is mostly sensitive to 
waveform errors $\delta h$ {\em tangential} to the signal manifold.  If the
errors in hybrid-waveforms are predominantly orthogonal to the signal manifold, then Eq.~(\ref{eq:Q-discussion}) may be overly restrictive.

In addition, let us briefly summarize how our findings impact current
efforts in GW detection.  The Ninja-2 project
\cite{NinjaWebPage,ninjashort} is well underway.  This project has
chosen fairly loose accuracy requirements to encourage wide
participation, namely NR waveforms that must be hybridized at a
frequency $M\omega_m<0.075$, with phase accuracy of the numerical
waveform better than $0.5$ rad.  This phase accuracy seems sufficient
for event {\em detection} purposes
(cf. Figs.~\ref{fig:Numerical_Resolution}
and~\ref{fig:SystematicErrors}), but matching at such high frequencies
will be dominated by PN errors to a degree that would likely impact
parameter estimation, (cf. Fig.~\ref{fig:T1vsT3vsT4_matching}).  While the Ninja-2 project focuses primarily on event detection, some of its efforts are geared at parameter estimation.  These efforts will be most useful in shedding light on the accuracy requirements discussed in this paper.  Future
Ninja projects may have to sharpen their accuracy requirements to
avoid biasing the results by inadequate model waveforms and will likely be targeted to improve understanding of parameter estimation.

This work can of course be extended in several
directions.  Most immediately, our conclusions are limited by the length of the employed NR
waveform, despite it being the longest published NR simulation.
Therefore, it will be useful to revisit the equal-mass, non-spinning
case when longer numerical data become available.  Furthermore, the
study needs to be extended to BBHs with different
mass ratios and spins.  With increasing mass ratio, the binary spends
more orbits in the strong-field regime (inversely proportional to the
symmetric mass ratio), so it is likely that unequal mass binaries will
require yet longer numerical simulations than the equal-mass case
considered here, as also pointed out by~\cite{Damour:2010}.  Spin
effects are
known only to lower PN order, so it is likely that
black hole spin may necessitate longer NR simulations
as well, dependent on their initial spin orientation.

Besides these obvious extensions based on refined numerical
simulations, there are also conceptual questions that deserve further
attention.  Eq.~(\ref{eq:Q-discussion}) incorporates
a detector network only in the crudest possible way.  It would be interesting to
investigate in more detail how Eq.~(\ref{eq:Q-discussion}) generalizes to a network
of multiple detectors.  
More detailed investigations into accuracy standards would also be valuable. On the one hand, Eq.~(\ref{eq:Q-discussion}) is a  sufficient condition and necessary accuracy standards for parameter
  estimation may be considerably weaker.  On the other hand, even when a certain error $\delta
h$ satisfies Eq.~(\ref{eq:Q-discussion}), it might nevertheless lead
to systematic bias in parameter estimation ~\cite{CutlerV:2007}.  One must also consider
how the presence of calibration errors affects the present
conclusions.  Finally, throughout this paper we have applied a time
and phase shift to minimize $||\delta h||$ between these two
waveforms.  This is generally acceptable, but doing so discards
information about the time of merger and phase of merger of
the binary.  A separate study will be necessary to determine how to
measure the binary's time of merger as accurately as possible, for
instance, in studies of electromagnetic counterparts.


 \ack We would like to thank Michael Boyle and Nick Taylor for
 providing the NR waveforms that we are analyzing in this paper. We
 also thank Tanja Hinderer, Lucia Santamar\'ia and Michele Vallisneri
 for careful reading of this manuscript.  It is a
 pleasure to acknowledge useful discussions with P. Ajith, Mike Boyle, 
Duncan
 Brown, Alessandra Buonnano, Curt Cutler, Scott Hughes, Lee Lindblom, Frank Ohme, Ben Owen, and Larne
 Pekowski.  H.P. gratefully acknowledges support from the
 NSERC of Canada, from the Canada Research Chairs Program, and from
 the Canadian Institute for Advanced Research. SMN's research was
 carried out at the Jet Propulsion Laboratory and the California
 Institute of Technology, under a contract with
 the National Aeronautics and Space Administration. \\



\bibliographystyle{unsrt}
\bibliography{References/References}

\begin{thebibliography}{10}

\bibitem{Barish:1999}
Barry~C. Barish and Rainer Weiss.
\newblock {LIGO} and the detection of gravitational waves.
\newblock {\em Phys. Today}, 52(10):44--50, Oct 1999.

\bibitem{Sigg:2008}
Daniel Sigg and the LIGO Scientific~Collaboration.
\newblock Status of the {LIGO} detectors.
\newblock {\em Class.\ Quantum Grav.}, 25(11):114041, 2008.

\bibitem{Acernese:2008}
F.~Acernese et~al.
\newblock Virgo status.
\newblock {\em Class.\ Quantum Grav.}, 25(18):184001, 2008.

\bibitem{Kuroda:2010}
K~Kuroda and the LCGT~Collaboration.
\newblock Status of {LCGT}.
\newblock {\em Class.\ Quantum Grav.}, 27(8):084004, 2010.

\bibitem{lisa}
{{Prince}, T.~A., {Binetruy}, P., {Centrella}, J., {Finn}, L.~S., {Hogan}, C.
  and {Nelemans}, G., {Phinney}, E.~S., {Schutz}, B. and LISA ~International
  ~Science ~Team}.
\newblock {LISA: probing the Universe with gravitational waves}.
\newblock Technical report, {LISA science case document}, 2007.
\newblock Available as \url{http://list.caltech.edu/mission_documents}.

\bibitem{lisa98}
P.~L. {Bender} et~al.
\newblock {LISA} - {L}aser interferometer space antenna for the detection and
  observation of gravitational waves.
\newblock Technical report, Garching, Germany, 1998.

\bibitem{Jennrich:2009}
O~Jennrich.
\newblock {LISA} technology and instrumentation.
\newblock {\em Class.\ Quantum Grav.}, 26:153001, Aug 2009.

\bibitem{AbadieLSC:2010}
J.~Abadie et~al.
\newblock {Predictions for the Rates of Compact Binary Coalescences Observable
  by Ground-based Gravitational-wave Detectors}.
\newblock {\em Class. Quant. Grav.}, 27:173001, 2010.
\newblock arXiv:1003.2480.

\bibitem{Blanchet2006}
Luc Blanchet.
\newblock Gravitational radiation from post-{N}ewtonian sources and
  inspiralling compact binaries.
\newblock {\em Living Rev.~Rel.}, 9(4), 2006.

\bibitem{Pretorius2007a}
Frans Pretorius.
\newblock Binary black hole coalescence.
\newblock In Monica Colpi, Piergiorgio Casella, Vittorio Gorini, Ugo Moschella,
  and Andrea Possenti, editors, {\em Physics of Relativistic Objects in Compact
  Binaries: From Birth to Coalescence}, volume 359 of {\em Astrophysics and
  Space Science Library}, pages 305--369. Springer Netherlands, 2009.
\newblock arXiv:0710.1338 [gr-qc].

\bibitem{Hinder:2010vn}
Ian Hinder.
\newblock {The Current Status of Binary Black Hole Simulations in Numerical
  Relativity}.
\newblock {\em Class. Quant. Grav.}, 27:114004, 2010.

\bibitem{Centrella:2010}
J.~M.. {Centrella}, J.~G. {Baker}, B.~J. {Kelly}, and J.~R. {van Meter}.
\newblock {Black-hole binaries, gravitational waves, and numerical relativity}.
\newblock {\em Rev.\ Mod.\ Phys.}, 82:3069, 2010.

\bibitem{Ajith:2008b}
P.~Ajith, S.~Babak, Y.~Chen, M.~Hewitson, B.~Krishnan, A.~M. Sintes, J.~T.
  Whelan, B.~Br\"ugmann, P.~Diener, N.~Dorband, J.~Gonzalez, M.~Hannam,
  S.~Husa, D.~Pollney, L.~Rezzolla, L.~Santamar\'\i{}a, U.~Sperhake, and
  J.~Thornburg.
\newblock Erratum: Template bank for gravitational waveforms from coalescing
  binary black holes: Nonspinning binaries [phys. rev. d 77, 104017 (2008)].
\newblock {\em Phys. Rev. D}, 79(12):129901, Jun 2009.

\bibitem{Santamaria:2010yb}
L.~Santamar\'{i}a, F.~Ohme, P.~Ajith, B.~Br{\"u}gmann, N.~Dorband, M.~Hannam,
  S.~Husa, P.~M{\"o}sta, D.~Pollney, C.~Reisswig, E.~L. Robinson, J.~Seiler,
  and B.~Krishnan.
\newblock Matching post-newtonian and numerical relativity waveforms:
  {S}ystematic errors and a new phenomenological model for non-precessing black
  hole binaries.
\newblock {\em Phys.\ Rev.\ D}, 82:064016, 2010.

\bibitem{NinjaWebPage}
The {NINJA} collaboration.
\newblock www.ninja-project.org.

\bibitem{ninjashort}
B.~Aylott et~al.
\newblock {Testing gravitational-wave searches with numerical relativity
  waveforms: Results from the first Numerical INJection Analysis (NINJA)
  project}.
\newblock {\em Class.\ Quantum Grav.}, 26(16):165008, 2009.

\bibitem{NRARwebsite}
The numerical relativity and analytical relativity ({NRAR}) collaboration.
\newblock https://www.ninja-project.org/doku.php?id=nrar:home.

\bibitem{Buonanno2007}
Alessandra Buonanno, Yi~Pan, John~G. Baker, Joan Centrella, Bernard~J. Kelly,
  Sean~T. McWilliams, and James~R. van Meter.
\newblock Approaching faithful templates for nonspinning binary black holes
  using the effective-one-body approach.
\newblock {\em Phys.\ Rev.\ D}, 76:104049, 2007.

\bibitem{Buonanno:2009qa}
Alessandra Buonanno, Yi~Pan, Harald~P. Pfeiffer, Mark~A. Scheel, Luisa~T.
  Buchman, and Lawrence~E. Kidder.
\newblock {Effective-one-body waveforms calibrated to numerical relativity
  simulations: coalescence of non-spinning, equal- mass black holes}.
\newblock {\em Phys.\ Rev.\ D}, 79:124028, 2009.

\bibitem{Buonanno99}
Alessandra Buonanno and Thibault Damour.
\newblock Effective one-body approach to general relativistic two-body
  dynamics.
\newblock {\em Phys. Rev. D}, 59(8):084006, 1999.

\bibitem{Miller2005}
Mark~A. Miller.
\newblock Accuracy requirements for the calculation of gravitational waveforms
  from coalescing compact binaries in numerical relativity.
\newblock {\em Phys.\ Rev.\ D}, 71:104016, 2005.

\bibitem{Lindblom2008}
Lee Lindblom, Benjamin~J. Owen, and Duncan~A. Brown.
\newblock {Model Waveform Accuracy Standards for Gravitational Wave Data
  Analysis}.
\newblock {\em Phys. Rev. D}, 78:124020, 2008.

\bibitem{Lindblom2009a}
Lee Lindblom.
\newblock Optimal calibration accuracy for gravitational wave detectors.
\newblock {\em Phys.\ Rev.\ D}, 80:042005, 2009.

\bibitem{Lindblom2009b}
Lee Lindblom.
\newblock Use and abuse of the model waveform accuracy standards.
\newblock {\em Phys.\ Rev.\ D}, 80(6):064019, 2009.

\bibitem{Lindblom:2010mh}
Lee Lindblom, John~G. Baker, and Benjamin~J. Owen.
\newblock {Improved Time-Domain Accuracy Standards for Model Gravitational
  Waveforms}.
\newblock {\em Phys. Rev.}, D82:084020, 2010.

\bibitem{Buonanno:2009}
Alessandra Buonanno, Bala~R. Iyer, Evan Ochsner, Yi~Pan, and B.~S.
  Sathyaprakash.
\newblock Comparison of post-newtonian templates for compact binary inspiral
  signals in gravitational-wave detectors.
\newblock {\em Phys. Rev. D}, 80(8):084043, Oct 2009.

\bibitem{Hannam:2010}
M.~{Hannam}, S.~{Husa}, F.~{Ohme}, and P.~{Ajith}.
\newblock {Length requirements for numerical-relativity waveforms}.
\newblock {\em Phys. Rev. D}, 82(12):124052, 2010.

\bibitem{Damour:2010}
T.~{Damour}, A.~{Nagar}, and M.~{Trias}.
\newblock {Accuracy and effectualness of closed-form, frequency-domain
  waveforms for nonspinning black hole binaries}.
\newblock {\em Phys. Rev. D}, 83(2):024006, 2011.

\bibitem{Damour:2000zb}
Thibault Damour, Bala~R. Iyer, and B.~S. Sathyaprakash.
\newblock Comparison of search templates for gravitational waves from binary
  inspiral.
\newblock {\em Phys.\ Rev.\ D}, 63(4):044023, Jan 2001.

\bibitem{Boyle:2011dy}
Michael Boyle.
\newblock {The uncertainty in hybrid gravitational waveforms: Optimizing
  initial orbital frequencies for binary black-hole simulations}.
\newblock {\em Phys. Rev. D (submitted)}, 2011.

\bibitem{LauPfeiffer2008}
Stephen~R. Lau, Harald~P. Pfeiffer, and Jan~S. Hesthaven.
\newblock {IMEX} evolution of scalar fields on curved backgrounds.
\newblock {\em Commun. Comput. Phys.}, 6:1063--1094, 2009.

\bibitem{Hennig:2008af}
Jorg Hennig and Marcus Ansorg.
\newblock {A Fully Pseudospectral Scheme for Solving Singular Hyperbolic
  Equations}.
\newblock {\em J. Hyperbol. Diff. Equat.}, 6:161, 2009.

\bibitem{LauLovelacePfeiffer2011}
Stephen~R. Lau, Geoffrey Lovelace, and Harald~P. Pfeiffer.
\newblock Implicit-explicit ({IMEX}) evolution of single black holes.
\newblock {\em In preparation}, 2011.

\bibitem{Nissanke2006}
Samaya Nissanke.
\newblock Post-{N}ewtonian freely specifiable initial data for binary black
  holes in numerical relativity.
\newblock {\em Phys.\ Rev.\ D}, 73:124002, 2006.

\bibitem{hannamEtAl:2007}
M~Hannam, S~Husa, B~Br{\"{u}}gmann, J~Gonz{\'{a}}lez, and U~Sperhake.
\newblock Beyond the {B}owen-{Y}ork extrinsic curvature for spinning black
  holes.
\newblock {\em Class.\ Quantum Grav.}, 24:S15, 2007.

\bibitem{Lovelace2009}
Geoffrey Lovelace.
\newblock Reducing spurious gravitational radiation in binary-black-hole
  simulations by using conformally curved initial data.
\newblock {\em Class.\ Quantum Grav.}, 26:114002, 2009.

\bibitem{JohnsonMcDaniel:2009dq}
Nathan~K. Johnson-McDaniel, Nicolas Yunes, Wolfgang Tichy, and Benjamin~J.
  Owen.
\newblock {Conformally curved binary black hole initial data including tidal
  deformations and outgoing radiation}.
\newblock {\em Phys.Rev.}, D80:124039, 2009.

\bibitem{Finn:1992}
Lee~S. Finn.
\newblock Detection, measurement, and gravitational radiation.
\newblock {\em Phys. Rev. D}, 46(12):5236--5249, Dec 1992.

\bibitem{CutlerFlanagan1994}
Curt Cutler and Eanna~E. Flanagan.
\newblock Gravitational waves from merging compact binaries: How accurately can
  one extract the binary's parameters from the inspiral waveform?
\newblock {\em Phys.\ Rev.\ D}, 49:2658--2697, 1994.

\bibitem{Shoemaker2009}
David Shoemaker.
\newblock Advanced {LIGO} anticipated sensitivity curves.
\newblock \url{https://dcc.ligo.org/cgi-bin/DocDB/ShowDocument?docid=2974},
  2009.
\newblock {L}IGO Document T0900288-v3.

\bibitem{Owen:1996}
B.~J. {Owen}.
\newblock {Search templates for gravitational waves from inspiraling binaries:
  Choice of template spacing}.
\newblock {\em Phys. Rev. D}, 53:6749--6761, June 1996.

\bibitem{McWilliams2010b}
Sean~T. McWilliams, Bernard~J. Kelly, and John~G. Baker.
\newblock Observing mergers of nonspinning black-hole binaries.
\newblock {\em Phys.\ Rev.\ D}, 82:024014, 2010.

\bibitem{Hannam:2009hh}
Mark Hannam, Sascha Husa, John~G. Baker, Michael Boyle, Bernd Bruegmann, Tony
  Chu, Nils Dorband, Frank Herrmann, Ian Hinder, Bernard~J. Kelly, Lawrence~E.
  Kidder, Pablo Laguna, Keith~D. Matthews, James~R. van Meter, Harald~P.
  Pfeiffer, Denis Pollney, Christian Reisswig, Mark~A. Scheel, and Deirdre
  Shoemaker.
\newblock {The Samurai Project: verifying the consistency of black- hole-binary
  waveforms for gravitational-wave detection}.
\newblock {\em Phys.\ Rev.\ D}, 79:084025, 2009.

\bibitem{Arun:2009}
K.~G. Arun, Alessandra Buonanno, Guillaume Faye, and Evan Ochsner.
\newblock Higher-order spin effects in the amplitude and phase of gravitational
  waveforms emitted by inspiraling compact binaries: Ready-to-use gravitational
  waveforms.
\newblock {\em Phys. Rev. D}, 79(10):104023, May 2009.

\bibitem{Kidder:2007rt}
Lawrence~E. Kidder.
\newblock {Using full information when computing modes of post-{N}ewtonian
  waveforms from inspiralling compact binaries in circular orbit}.
\newblock {\em Phys.\ Rev.\ D}, 77:044016, 2008.

\bibitem{Boyle2007}
Michael Boyle, Duncan~A. Brown, Lawrence~E. Kidder, Abdul~H. Mrou{\'e},
  Harald~P. Pfeiffer, Mark~A. Scheel, Gregory~B. Cook, and Saul~A. Teukolsky.
\newblock High-accuracy comparison of numerical relativity simulations with
  post-{N}ewtonian expansions.
\newblock {\em Phys.\ Rev.\ D}, 76:124038, 2007.

\bibitem{Damour:2002kr}
Thibault Damour, Bala~R. Iyer, and B.~S. Sathyaprakash.
\newblock Comparison of search templates for gravitational waves from binary
  inspiral: 3.5{PN} update.
\newblock {\em Phys.\ Rev.\ D}, 66(2):027502, Jul 2002.

\bibitem{Baker2006e}
John~G. Baker, Sean~T. McWilliams, James~R. van Meter, Joan Centrella, Dae-Il
  Choi, Bernard~J. Kelly, and Michael Koppitz.
\newblock Binary black hole late inspiral: Simulations for gravitational wave
  observations.
\newblock {\em Phys.\ Rev.\ D}, 75:124024, 2007.

\bibitem{Buonanno-Cook-Pretorius:2007}
Alessandra Buonanno, Gregory~B. Cook, and Frans Pretorius.
\newblock Inspiral, merger, and ring-down of equal-mass black-hole binaries.
\newblock {\em Phys.\ Rev.\ D}, 75(12):124018, 2007.

\bibitem{Hannam2007}
Mark Hannam, Sascha Husa, Jos{\'e}~A. Gonz{\'a}lez, Ulrich Sperhake, and Bernd
  Br{\"u}gmann.
\newblock Where post-{N}ewtonian and numerical-relativity waveforms meet.
\newblock {\em Phys.\ Rev.\ D}, 77:044020, 2008.

\bibitem{SpECwebsite}
\url{http://www.black-holes.org/SpEC.html}.

\bibitem{Scheel2008}
Mark~A. Scheel, Michael Boyle, Tony Chu, Lawrence~E. Kidder, Keith~D. Matthews,
  and Harald~P. Pfeiffer.
\newblock High-accuracy waveforms for binary black-hole inspiral, merger, and
  ringdown.
\newblock {\em Phys.\ Rev.\ D}, 79:024003, 2009.

\bibitem{Chu2009}
Tony Chu, Harald~P. Pfeiffer, and Mark~A. Scheel.
\newblock {High accuracy simulations of black hole binaries:spins anti-aligned
  with the orbital angular momentum}.
\newblock {\em Phys. Rev.}, D80:124051, 2009.

\bibitem{Szilagyi:2009qz}
Bela Szilagyi, Lee Lindblom, and Mark~A. Scheel.
\newblock {Simulations of Binary Black Hole Mergers Using Spectral Methods}.
\newblock {\em Phys.\ Rev.\ D}, 80:124010, 2009.

\bibitem{Blackman-Tukey}
R.B. Blackman and J.W. Tukey.
\newblock {\em The Measurement of Power Spectra, From the Point of View of
  Communications Engineering}.
\newblock Dover Books, New York, NY, first edition, 1959.

\bibitem{McKechan:2010kp}
D.J.A. McKechan, C.~Robinson, and B.S. Sathyaprakash.
\newblock {A tapering window for time-domain templates and simulated signals in
  the detection of gravitational waves from coalescing compact binaries}.
\newblock {\em Class.Quant.Grav.}, 27:084020, 2010.

\bibitem{Reisswig2009}
C.~{Reisswig}, N.~T. {Bishop}, D.~{Pollney}, and B.~{Szil{\'a}gyi}.
\newblock {Unambiguous Determination of Gravitational Waveforms from Binary
  Black Hole Mergers}.
\newblock {\em Physical Review Letters}, 103(22):221101, November 2009.

\bibitem{HannamEtAl:2010}
Mark Hannam, Sacha Husa, Franke Ohme, Doreen M\"{u}ller, and Bernd
  Br\"{u}gmann.
\newblock Simulations of black-hole binaries with unequal masses or
  nonprecessing spins: {A}ccuracy, physical properties, and comparison with
  post-{N}ewtonian results.
\newblock {\em Phys.\ Rev.\ D}, 82:124008, 2010.

\bibitem{Boyle-Mroue:2008}
{M. Boyle, A.H. Mrou\'e }.
\newblock Extrapolating gravitational-wave data from numerical simulations.
\newblock {\em Phys.\ Rev.\ D}, 80:124045, 2009.

\bibitem{Hannam2007c}
Mark Hannam, Sascha Husa, Bernd Br{\"u}gmann, and Achamveedu Gopakumar.
\newblock Comparison between numerical-relativity and post-{N}ewtonian
  waveforms from spinning binaries: the orbital hang-up case.
\newblock {\em Phys.\ Rev.\ D}, 78:104007, 2008.

\bibitem{MroueGpuTalk2010}
Abdul~H. Mrou\'e.
\newblock Binary black hole simulations using {CUDA}.
\newblock {NVIDIA} {GPU} Technology Conference, 2010.

\bibitem{DIN}
T.~{Damour}, B.~R. {Iyer}, and A.~{Nagar}.
\newblock {Improved resummation of post-Newtonian multipolar waveforms from
  circularized compact binaries}.
\newblock {\em Phys.\ Rev.\ D}, 79:064004, 2009.

\bibitem{CutlerV:2007}
C.~{Cutler} and M.~{Vallisneri}.
\newblock {LISA detections of massive black hole inspirals: Parameter
  extraction errors due to inaccurate template waveforms}.
\newblock {\em Phys. Rev. D}, 76(10):104018, November 2007.

\end{thebibliography}
\end{document}